\documentclass{JHEP}
\usepackage{graphicx,amsmath,amssymb}
\usepackage{epsfig,multicol}
\input{epsf.sty}
\newcommand{\ba}{\begin{eqnarray}}
\newcommand{\ea}{\end{eqnarray}}
\newcommand{\D}{\overline{\mbox{D}}}
\newcommand{\e}{\epsilon}
\newcommand{\Pm}{\Psi_{({\bf m})}}
\newcommand{\Em}{E_{\bf m}}
\newcommand{\bm}{\beta_{2} }
\newcommand{\Bm}{B_{({\bf m})}}
\newcommand{\bt}{\beta_{21} }
\newcommand{\br}{\beta_{23} }

\title{The Shape of Gravity in a Warped Deformed Conifold}
\author{Hassan Firouzjahi\footnote{Electronic mail:
firouzh@lepp.cornell.edu}
~and S.-H.~Henry Tye\footnote{Electronic mail:
tye@lepp.cornell.edu}
\\Newman Laboratory for Elementary Particle Physics, Cornell University, 
Ithaca, NY 14853}
\vspace{0.5cm}

\abstract{We study the spectrum of the gravitational modes in Minkowski spacetime due to a 6-dimensional warped deformed conifold, i.e., a warped throat, in superstring theory. After identifying the zero mode as the usual 4D graviton, we present the KK spectrum as well as other excitation modes. Gluing the throat to the bulk (a realistic scenario), we see that the graviton has a rather uniform probability distribution everywhere while a KK mode is peaked in the throat, as expected. 
Due to the suppressed measure of the throat in the wave function normalization,
we find that a KK mode's probability in the bulk can be comparable to that of the graviton mode.
We also present the tunneling probabilities of a KK mode from the inflationary throat to the bulk and to another throat. Due to resonance effect, the latter may not be suppressed as natively expected. 
Implication of this property to reheating after brane inflation is discussed.
}

\begin{document}

\maketitle

\section{Introduction}

In flux compactification in 10-dimensional superstring theory \cite{Giddings:2001yu}, 
such as a KKLT vacuum \cite{Kachru:2003aw},
all moduli of the 6 dimensional compact manifold are stabilized. 
Presumably, our today's universe is described by such a KKLT-like vacuum. 
Such a manifold typically includes warped throats inside the bulk of the manifold.
For a 4-dimensional Minkowski spacetime, a massless 4D graviton is always present as a zero mode.
In addition, there are excitation modes including Kaluza-Klein (KK) modes.
It is possible that we (i.e., the standard model particles) live inside a stack of 
${\D}$3-branes sitting at the bottom of such a throat. So it will be interesting to know the 
spectrum of the closed string modes in the throat. The spectrum and their respective wave functions
will reveal the structure of the throat as well as the bulk.

Another reason in understanding this spectrum and their respective wave functions has to do 
with the reheating after inflation. 
It is generally believed that inflation explains the origin of our universe.
A concrete realization of the inflationary scenario in superstring theory is the D-${\D}$-brane
inflationary scenario \cite{Dvali:1998pa,collection}. In the KKLMMT scenario \cite{Kachru:2003sx}, where the 6 compactified dimensions are dynamically stabilized, 
brane inflation is realized by introducing an extra D3-${\D}$3-brane pair in a KKLT-like vacuum.
The $\D$3-brane is sitting at the bottom of a Klebanov-Strassler (KS) type throat \cite{Klebanov:2000hb}, 
as the mobile D3-brane moves towards it. 
This describes the inflationary epoch, where the vacuum energy driving inflation comes from the
extra D3-${\D}$3-brane tension and a possible conformal coupling term. 
In this scenario, inflation ends when the D3-${\D}$3-branes collide and annihilate, initiating 
the hot big bang epoch. If this annihilation happens in a throat (A throat) other than the standard model 
throat (S throat), one likes to ask if the resulting reheating is compatible with big bang nucleosynthesis. 
This issue has been studied recently \cite{Barnaby:2004gg,Kofman:2005yz,Chialva:2005zy,Frey:2005jk}. A key factor involves the tunneling rate, i.e., the probability amplitude of a KK mode (which is produced in the A throat) in the bulk and the S throat. Here we examine in more details the KK wave functions.

We start with the KS throat. 
Given the metric, it is quite straightforward to solve the Schr\"{o}dinger-like equation to
obtain the KK spectrum of the 10D graviton. We shall assume 4D Minkowski spacetime to be a good approximation to the KKLT-like vacuum. We find a zero mode to be identified as the massless 4D graviton, while the discrete KK mass spectrum $\Em$ is given by solving a Schr\"{o}dinger-like equation. 
In the $S^{3} \times S^{2}$ approximation around the bottom 
of the throat, the mass eigenvalues may be classified by the corresponding quantum numbers
($j-1$) for $S^{3}$ and $l$ for $S^{2}$.
It is convenient to introduce a dimensionless ${\hat E}_{\bf m}$, $\Em \sim {\hat E}_{\bf m} {h_{A}}/{l_{s}}$
where $h_{A} <1$ is the warp factor at the bottom of the A throat and $l_{s}$ is the string scale. 
In a WKB approximation, which is accurate to around a few percent, we find that
${\hat E}_{\bf m}$ for ${\bf m}=(n, j-1, l)$ (where $j-1$ and $l$ are the $S^{3} \times S^{2}$ quantum numbers) is given by
\ba
\sqrt{{\hat E}_{\bf m}^2-{\cal{Q}}^2}-
{\cal{Q}}\,\arctan \left(\frac{\sqrt{  {\hat E}_{\bf m}^2-{\cal{Q}}^2}}{{\cal{Q}}}\right)
=1.03\, (n-\frac{1}{2} + \frac{\delta_{0,l}}{4}) \nonumber \\
{\cal{Q}}^2\equiv 1.4\, (j^2-1) +3.4\, l(l+1) + .65\, .
\ea
and the numerical values are chosen to get a good fit to the actual spectrum.
The lowest  is then ${\hat E}_{\bf 1}=2.02 $, which,  for slow-roll KKLMMT brane inflation 
\cite{Firouzjahi:2005dh},  takes value of order $ \Em \sim 10^{-4} M_{P}$.

Next we like to look at the localization of the KK wavefunctions and their tunneling properties. We may restrict ourselves to the S-waves.
For $ j -1= l = 0$, we have 
\ba
{\hat E}_{{\bf m}=(n,0,0)} = {\hat E}_{n}=1.04\, n +1.01   
\ea
We note that both the eigenvalues and eigenvectors are well approximated by Bessel functions. 
The eigenfunctions are given by Bessel functions $J_{\nu}$ of 
order $\nu \simeq 2.44$, but with an unusual argument, while the eigenvalues are related to, not surprisingly, the zeros of $J_{\nu}$.

Next we consider the more realistic scenario where the manifold is made of a bulk with a number of throats. Treating the bulk crudely, continuity requires the matching of each wavefunction and its derivative at the edge of the throat. Now, an excitation eigenfunction becomes a linear combination of  
$J_{\nu}$  and $Y_{\nu}$ in the throat and is approximated by sinusoidal functions in the bulk. 
The wave function of the $n$th KK mode is peaked at the bottom of the throat, as expected. The 
ratio of the wave function $\Psi_{n}(\tau_{c})$ at the edge of the throat (where the warp factor 
$h_{A}(\tau_{c}) \simeq 1$, so $\Psi_{n}(\tau_{c})$ is comparable to the wave function in the bulk) to that at the bottom of the throat $\Psi_{n}(0)$ with warp factor $h_{A}=h_{A}(0)$ is given by
\ba
\left| \frac{\Psi_{n}(\tau_{c})}{\Psi_{n}(0)} \right|^{2}  \simeq E_{n}^{8} \sim ((n+1) h_{A})^{8}
\ea
where the KK mass $E_{n} \simeq {\hat E}_{n} h_{A} \simeq (n +1) h_{A}$ (in string units). 
Another interesting quantity is the ratio of $\Psi_{n}(\mbox{bulk})
\simeq \Psi_{n}(\tau_{c})$ to that of the graviton mode $\Psi_{g}(\mbox{bulk})$ at the edge of the throat. 
Since the graviton is everywhere, this provides a measure of how much the $n$th KK mode can get out of the throat when the time scale is longer than the tunneling time but shorter than the KK mode lifetime.
We find that there are two limiting situations. In the case where the manifold is made up of a set of throats only, 
 \ba
\left|\frac{\Psi_{n}(\mbox{bulk})}{\Psi_{g} (\mbox{bulk})} \right|^{2}  \sim E_{n}^{6} \sim ((n+1)h_{A})^{6}
\ea
This suggests that the tunneling from one throat to another is very much suppressed. However, in the case where the manifold has a reasonable bulk (that is, a region where there is little or no warping), then
 \ba
\left|\frac{\Psi_{n}(\mbox{bulk})}{\Psi_{g}(\mbox{bulk})} \right|^{2}  \sim 1
\ea
That is, the average KK wavefunction in the bulk is essentially the same as the average graviton wavefunction. This result is due to the fact that the warped metric weights the throat significantly less than the bulk in the measurement of the volume of the compactified manifold, so the wave function normalization pushes the wave function to the bulk. 

On the other hand, the relevant tunneling probability for reheating may be the transmission coefficient, which measures the tunneling from the A throat to the bulk (and to another throat via the bulk), where there is only outgoing wave (i.e., away from the barrier) in the final stage. It is important to decide which of the estimates is relevant for reheating. Because the compactified manifold has finite size, the above ratio estimates may be more relevant if the KK mode decays slowly so its final reflection component should be included. If it decays fast enough before reflection, then the transmission coefficient should be more relevant. In this case, we find 
  \ba
 T(\mbox{ A} \rightarrow \mbox{bulk}) \simeq \frac{4}{\Theta_{A}^{2}}\sim (E_{n})^8 \sim ((n+1)\, h_A)^8
 \ea
 where $\Theta_{A}/2$ is the standard WKB integral of the (imaginary) momentum across the barrier.
 The transmission coefficient from the A throat to  the S throat via the bulk is given by
\ba
T (A \rightarrow S)=4\, \left( ( \Theta_{A}^2 + \frac{1}{\Theta_{A} ^2})^{2} 
\cos^{2} L + 4\, \sin^{2} L\right)^{-1}
\label{transm1}
\ea
where $E_{n}(A)=(n+1) h_{A}$ matches $E_{n'}(S) \simeq (n'+1) h_{S}$ in the S throat. In a realistic model, the KK modes have widths and this matching is not hard if the warp factor of the S throat
 $h_{S} \ll h_{A}$.
In the case where the manifold is made up of a set of throats only (mimicking the two throats model in the RS scenario as in \cite{Barnaby:2004gg, Kofman:2005yz,Chialva:2005zy}), so $L\simeq 0$,
we find that $T$ is extremely small,
\ba
T (A \rightarrow S) \sim (\Theta_A )^{-4} \sim ((n+1) h_{A})^{16}
\ea
One should compare this to that in the Randall-Sundrum (RS) scenario \cite{Randall:1999ee,Randall:1999vf}, 
where this ratio is found to go like $h_{A}^{4}$ \cite{Barnaby:2004gg, Kofman:2005yz,Chialva:2005zy}. 
This latter result follows from Ref.\cite{Dimopoulos:2001ui} where the $h_{A}^{4}$ comes simply from the warped metric.
There, the wave function obeys the Israel junction condition in the infrared region, which is very specific to the 5-dimensional spacetime in the RS model. Here we are dealing with a 
6-dimensional warped deformed conifold in 10 dimensional spacetime, where the wave function at the bottom of the throat must be finite. For $h_{A} \sim 1/5$ to $10^{-3}$ or smaller, this difference is huge.

However, for some KK modes, $L=(n_{L}+1/2) \pi$, so $\cos L =0$.
For these KK modes, the transmission coefficient (\ref{transm1}) approaches unity, 
\ba
T (A \rightarrow S)  \sim 1
\ea
This is the well-known resonance effect.
Based on a simple estimate, we expect that some of the KK modes will satisfy this condition.
So our two ways of estimating tunneling of KK modes from one throat to another agree qualitatively.
The end result is that the tunneling depends crucially on the presence/size of the bulk, bracketing the tunneling amplitude in the simpler RS scenario. 

The implication to reheating after inflation is interesting. In a realistic Type IIB compactified manifold with multi-throats, we expect the presence of a bulk.
In this scenario, the KK modes are produced in the inflationary throat (the A throat) where the mobile D3-brane falls into this throat, collides and annihilates with the $\D$3 brane sitting at the bottom of the throat. These KK modes will easily tunnel to another throat, say the S throat that contains the standard model branes. These KK modes can then decay into the open strings in the S throat. 
As pointed out in Ref.\cite{Barnaby:2004gg,Kofman:2005yz,Chialva:2005zy},
this will enable the excitation of the standard model modes and reheat the universe to start the hot big bang epoch, consistent with big bang nucleosynthesis. The differences of our result compared to that in  Ref.\cite{Barnaby:2004gg,Kofman:2005yz,Chialva:2005zy} are the two effects : \\
(1) the bound state property in a warped deformed conifold actually suppresses the tunneling much more than the 5-dimensional RS scenario; \\
(2) the presence of the (un-warped) bulk can provide a huge enhancement of the tunneling amplitude. 
 \\
These two effects tend to offset each other. In the supergravity approximation, the presence of a 
finite-size bulk is expected, so the second effect may win. In a realistic model, under appropriate conditions, we expect the tunneling amplitude can be big enough to allow a successful reheating of the universe after brane inflation in a multi-throat model. However, the answer may be quite sensitive to the details of the model.

\section{Setup and the Graviton}

We are interested in the 10D graviton spectrum. We assume that the background is a static
geometry with Poincare symmetry in four dimensions. Lacking a deep dynamical understanding,
a 4D Minkowski metric is achievable only with some fine-tuning.
There are three kinds of metric fluctuations, labeled by their spin from a 4D observer
point of view. They are scalar, vector and tensor modes. The scalar modes determine the size and shape of the the internal manifold (here Calabi-Yau manifold). The scalar modes which control the size of the compactification is known as radions.
The vector modes are fluctuations of metric in the form $g_{\mu\, m}$, i.e. metric field with one index in the extended four direction and one index in the internal dimension. 
The existence of these modes are model dependent. For example, the Randall-Sundrum (RS) model 
\cite{Randall:1999vf} has no massless vector mode due to the $Z_2$ symmetry.  
Finally, the graviton is the spin 2 fluctuation of the 4D metric, 
$\eta_{\mu\,\nu} \rightarrow \eta_{\mu\,\nu} +h_{\mu\,\nu} $. Beside the zero mode, the RS model has a tower of massive KK graviton. These modes are highly peaked in the IR region with the
mass gap $\delta m  \sim TeV$. 

Let us start by considering the following action
\ba
\label{action1}
S= M_s^{D-2}\int d^D\, x\, \sqrt{-g}\left( \frac{R}{2}+ {\cal L}   \right) + S_{topological}
\ea
where $M_s$ represents the D-mimensional mass scale.
We consider a general form of ${\cal L} $ such that a background solution with a 4D Minkowski metric
is possible.  ${\cal L} $ contains bulk contributions and it may also contain contributions from local sources.  For example ${\cal L} $ in the RS model contains a bulk
 cosmological constant and localized brane(s) tension(s). 
 In the GKP model, bsed on IIB string theory, ${\cal L} $ 
 has contribution from three form flux $G_{(3)}$, five flux ${\tilde{F}}_{(5)}$ and axio-dilaton field. 
 It also contains local source D3-branes, D7-branes and O-planes. 

The model admits the following metric
\ba
\label{ansatz2}
ds^2= {{g}}_{\mu\nu}(x,y)\, dx ^{\mu}\, dx^{\nu} + {g}_{mn}(y) dy^{m}\, dy^{n} \, .
\ea
where the compactified manifold (with the metric $g_{mn}$) is stabilize and remains static. 
 
The Einstein equation is 
 \ba
 \label{Ein1}
 R_{MN}= T_{MN}-\frac{1}{D-2} T g_{MN}
 \ea 
 where $M$ and $N$ runs on the entire space-time and $T$ is the trace of the stress energy tensor 
 $T_{M N}$, given by
 \ba
 T_{M N}= g_{MN}\,{\cal L}   -2 \frac{\delta {\cal L} }{\delta g^{MN}}
 \ea
 Suppose ${\cal L} $ depends only on the internal metric, i.e.,
 \ba
 \label{internal}
 \frac{\delta {\cal L} }{\delta g^{\mu\nu}}=0 \, .
 \ea
Intuitively this implies that the total contribution from integrating ${\cal L} $ over the internal dimensions behaves like a cosmological constant for a 4D observer. 
 Using the condition (\ref{internal}), the components of stress energy tensor are
\ba
T_{\mu\nu}=g_{\mu\nu} {\cal L}  \quad \quad ,\quad \quad 
 T_{m n}=g_{m n} {\cal L} + {\cal T}_{m n}
 \ea
 where ${\cal T}_{m n} \equiv -2 \delta {\cal L} / \delta g^{m n}$ \, .
 
 With the above form of the stress energy tensor and with the definition ${\cal T}={\cal T}^m_m$,
 the $\mu\nu$ components of the Einstein equation obey 
 \ba
 \label{Ein2}
 R_{\mu\nu}= - \frac{2 {\cal L}+{\cal T}}{D-2} \, g_{\mu\nu} \, .
 \ea 
It should be emphasized that the above compact form of the Einstein equation is true
everywhere, in the bulk or at the position of the local sources. At local sources like branes, there
would be a delta function singularity  for $R_{\mu\nu},  {\cal L} $ and ${\cal T}$ such that 
Eq.(\ref{Ein2}) is satisfied. 

With the metric (\ref{ansatz2}), $R_{\mu\nu}$ is given by
\ba
\label{Rim1}
R_{\mu\nu}={R^{(4)}}_{\mu\nu} -\frac{1}{2} {\nabla^2_{(D-4)}} {{g}}_{\mu\nu}
-\frac{1}{4} {{g}}^{\rho\lambda}\partial_{m}{{g}}_{\mu\nu}{\partial}^{m}{{g}}_{\rho\lambda}
+\frac{1}{2} {{g}}^{\rho\lambda}\partial_{m}{{g}}_{\nu\rho}{\partial}^{m}{{g}}_{\mu\lambda} \, .
\ea
Here ${R^{(4)}}_{\mu\nu} $ is the 4D Ricci tensor calculated purely from ${{g}}_{\mu\nu}$ while 
${\nabla^2_{(D-4)}}$ is the Laplacian constructed from ${g}_{mn} $.
 
For $g_{\mu\nu}(x,y) = e^{2A(y)} {\hat  g}_{\mu\nu}$,
where ${\hat  g}_{\mu\nu}$ is Ricci flat ($R^{(4)}_{\mu\nu} ({\hat g}) =0$),
Eq.(\ref{Ein2}) reduces to
\ba
\label{back}
\frac{2 {\cal L}+{\cal T}}{D-2}=\frac{1}{2} e^{-2A} \left({\nabla^2_{(D-4)}} e^{2A} 
+e^{-2A} \partial_{m} e^{2A}\,\partial^{m} e^{2A} \right)
\ea

The case we are interested here is 
\ba
\label{grav1}
g_{\mu\nu}= e^{2A(y)}(\eta_{\mu\nu} + h_{\mu\nu}(x^{\rho}, y^{m})) \, 
\ea
For an extensive analysis of other perturbations, see Ref. \cite{Giddings:2005ff}.

In the absence of the perturbation $h_{\mu\nu}$ and with the Minkowski background $\eta_{\mu\nu}$,
$A(y)$ satisfies Eq.(\ref{back}). The Minkowski metric implies effective zero 4D cosmological constant. In view of the compactification dynamics, this requires a fine tuning. The actual fine-tuning  required is for a very small positive 4D cosmological constant compatible with the observed dark energy. In this case, the Minkowski metric is a close enough approximation for our purpose.  
Next we like to solve for $h_{\mu\nu}$.

Imposing the transverse-traceless gauge conditions 
\ba
\label{TT}
h^{\mu}_{\mu}=\partial_{\mu} h^{\mu\nu} =0 \, .
\ea
we have
\ba
\label{deltaR}
\delta R_{\mu\nu}=-\frac{1}{2} \nabla^2_{(4)} h_{\mu\nu} -\frac{1}{2} \nabla^2_{(D-4)}(e^{2A} 
h_{\mu\nu} )-\frac{h_{\mu\nu}}{2} \partial_{m} e^{2A}\,\partial^{m} e^{2A} \, .
\ea
where $\nabla^2_{(4)}$ is constructed purely from $g_{\mu\nu}$.

Using this expression for $\delta R_{\mu\nu}$ and the relation imposed from the background solution 
Eq.(\ref{back}) in Eq(\ref{Ein2}), we obtain 
\ba
\label{eom1}
e^{-2A}\nabla^2_{(D-4)}(e^{2A} h_{\mu\nu}) -e^{-2A} h_{\mu\nu} \nabla^2_{(D-4)}(e^{2A} )+
\nabla^2_{(4)} h_{\mu\nu} =\nabla^2_{(D)} h_{\mu\nu} =0 \, .
\ea
Interestingly enough this is  the equation of motion for a D-dimensional massless
scalar field which was also noticed in \cite{Brandhuber:1999hb}. This makes sense because the gravitons in D-dimensions are massless
spin two particles and since we only excite the $\mu\nu$ components they behave like  
scalar fields from the internal dimension point of view. 
This equation is true everywhere in the bulk and the contributions of the local sources are included automatically. To obtain the boundary conditions at the positions of local sources, one simply integrate
$\nabla^2_{(D)} h_{\mu\nu}$ passing across the local source.  More specifically, in a Gauss
normal coordinate locally chosen near the source at position $y=y_0$, the boundary condition
obtained from Eq.(\ref{eom1}) is 
\ba
\label{bc}
\partial_{n_i}\, h_{\mu\nu} |^{y_0+\epsilon} _{y_0-\epsilon} = 0
\ea
where $\partial_{n_i}$ is the normal derivative along $x^i$ coordinate perpendicular to the source where $i$ runs from 1 to the number of transverse dimensions to the source.

The KK spectrum of the graviton is obtained from Eq.(\ref{eom1}), 
\ba
\label{eom2}
\nabla^2_{(D-4)}(e^{2A} h_{\mu\nu}) - h_{\mu\nu} \nabla^2_{(D-4)}(e^{2A} )+
{E_{\bf m}}^2 \, e^{2A} h_{\mu\nu} =0
\ea
where 
\ba
\label{hatE} 
{E_{\bf m}}^2\, h_{\mu\nu} = \eta^{\mu\nu}\partial_{\mu} \partial_{\nu}h_{\mu\nu} 
\ea
such that $E_{\bf m}$ is the mass
measured by a 4D observer.

It is clear from above equation (\ref{eom2}) that there always exists a zero mode with solution
$h_{\mu\nu}=$ constant. 
This represents the ordinary 4D graviton propagating in the four dimension with Poincare symmetry. Since the internal dimensions are compactified, the zero mode is normalizable. 
This in turn ensures that the effective four-dimensional gravity strength is finite. 
More explicitly, consider the decomposition of the graviton excitations in the following way
\ba
\label{decompose}
h_{\mu\nu}(x^{\rho},y^n)=\sum_{{\bf m}} h^{({\bf m})}_{\mu\nu} (x^{\rho})\,\Pm(y^n)
\ea
The gravitational part is of the action (\ref{action1})
\ba
\label{action2}
S_g &=&\frac{M_s^{D-2}}{2} \int d^D\, x\, \sqrt{-g}\, R \nonumber\\ 
&\sim& \frac{M_s^{D-2}}{2} \int d^4\, x\, d^{D-4} y\, h^{\mu\nu}\,  \partial^2
h_{\mu\nu}  e^{2A}\, \sqrt{|g_{a b}|}\nonumber\\
&=&\frac{M_s^{D-2}}{2} \int d^4\, x \sum_{{\bf m,m'}} h^{(m)\,\mu\nu} h^{(m')}_{\mu\nu}
\int  d^{D-4}y \sqrt{|g_{ab}|} \, e^{2A} \, \Pm(y)\,\Psi_{({\bf m'})}(y)
\ea

The natural normalization is 
\ba
\label{ortho}
M_s^{D-2}\,\int  d^{D-4}y\, \sqrt{|g_{ab}|} \, e^{2A} \, \Pm(y)\,\Psi_{({\bf m'})}(y) =M_P^2\, \delta_{{\bf mm' }}
\ea
For the zero mode, $\Psi_{g}$,  we obtain the usual relation between the gravity strength in four dimensions and the fundamental mass scale of the higher dimensional theory.


After this general discussions we would like to find the KK spectrum of the gravitons and other closed string modes in the KS background. To do so, let us first review the geometry of a warped deformed conifold.

\begin{figure}[t]
\vspace{1cm}
   \centering
   \hspace{-1cm}\includegraphics[width=4in]{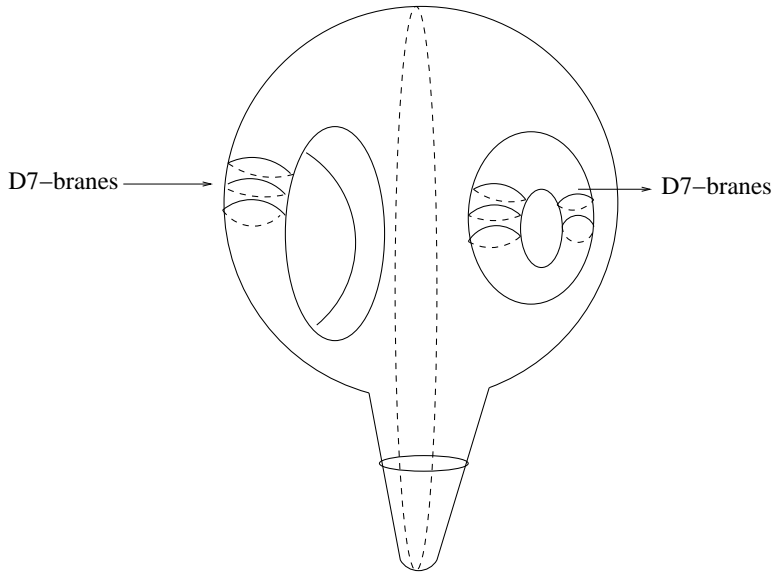}
   \hspace{1cm} 
\caption{A schematic picture of the Calabi-Yau manifold is presented here.
The large circle given by dashed line represent the 3-cycle 
where NS-NS  three form
$H_3$ is turned on. The smaller circle in the throat stands for the 3-cycle 
where the R-R three form $F_3$ is turned on. Also shown are D7-branes 
wraping 4-cycles. 
There may exists a number of throats 
like the one shown here. There is a mirror image of the entire picture 
due to the ${\bf{IIB/Z_2}}$ orientifold operation.
\label{CY}
}
\end{figure}

\section{A Throat in the Calabi-Yau Manifold}

A KKLT vacuum involves a Calabi-Yau (CY) manifold with fluxes \cite{Giddings:2001yu}.
Consider F-theory compactified on an elliptic CY 4-fold $X$. The F-theory 4-fold is a useful way of encoding the data of a solution of type IIB string theory; the base manifold M of the fibration encodes the IIB geometry, while the variation of the complex structure of the elliptic fiber describes the profile of the IIB axion-dilaton. In such a model, one has a tadpole condition on the Euler number $\chi$ of $X$
\ba
\label{euler}
\frac{\chi (X)}{ 24} = N_{D3} + \frac{1}{2\kappa^2_{10} T_3} 
\int_M H_3 \wedge F_3 
\ea
Here $T_3$ is the tension of a D3 brane, $N_{D3}$ is the net number of (D3 - ${\D}$3) branes  filling the noncompact dimensions, and $H_3$, $F_3$ are the 3-form fluxes in the IIB theory which arise in the NS and RR sector, respectively. In the absence of flux, it is always possible to deform such an F-theory model to a locus in moduli space where it can be thought of as an orientifold of a IIB CY compactification. So one may use the language of IIB orientifolds, with M being the CY 3-fold which is orientifolded. In this language, the LHS of Eq.(\ref{euler}) counts the negative D3-brane charge coming from the O3 planes and the induced D3 charge on D7 branes, while the terms on the RHS count the net D3 charge from transverse branes and fluxes in the CY manifold. The K\"{a}hler moduli are stabilized by non-perturbative dynamics.

 As shown in Figure \ref{CY}, there is a Klebanov-Strassler throat in the manifold \cite{Klebanov:2000hb}. The throat is a warped deformed conifolds, which is a non-compact CY 3-fold. This throat is glued to the compact CY manifold.  A number of such throats can be present in a typical compact CY manifold, although our discussion will focus on a specific throat. Such a CY manifold yields a KKLT-like vacuum with a tiny positive cosmological constant to explain the observed dark energy. It is expected to be metastable, with a lifetime much longer than the age of the universe.

During inflation, there is an additional D3 - ${\D}$3-brane pair (without changing $N_{D3}$ or any other term in Eq.(\ref{euler})). The ${\D}$3-brane is attracted to a throat, so it is expected to sit at the bottom of such a throat. On the other hand, the D3-brane is mobile. As the D3-brane moves towards the ${\D}$3-brane (due to the attractive force between them), inflation takes place (due to the extra tensions of this additional D3 - ${\D}$3-brane pair). At the end of inflation, the D3-brane collides and annihilates with the ${\D}$3-brane. As D3-${\D}$3 annihilation takes place, the universe is reheated and cosmic superstrings are produced.

A warped throat typically has a non-trivial geometry. The 6-dimensional
 metric around the throat may be written as
\ba
\label{metric5}
ds_6^2 = dr^2 +r^2ds_5^2
\ea
where the non-trivial 5-dimensional metric $ds_5^2$ may take a variety
of geometries. The best known example is $S^5$, which has $N=4$ supersymmetry. However, we are interested in $N=1$ supersymmetry, that is, base manifolds that yield CY 3-folds. 
To be specific, we shall focus on the simplest base manifold that has $N=1$ supersymmetry, namely, $T^{1,1}$ \cite{Romans:1984an,Klebanov:1998hh}. 
In fact, $T^{1,1}$ is the first known Sasaki-Einstein metric for $N=1$ supersymmetry, which is preserved 
by its deformation. 
This conifold and its deformation are well-studied 
\cite{Candelas:1989js,Minasian:1999tt,Ohta:1999we,Herzog:2001xk}.

\subsection{The Conifold}

A cone is defined by the following equation in ${\cal{\bf{C^4}}}$
\ba
\label{6cone}
\sum_{i=1}^{4} w_i^2=0
\ea
Here Eq.(\ref{6cone}) describes a smooth surface apart from the point $w_i=0$. 
The geometry around the conifold is studied in Ref.\cite{Candelas:1989js}.
If $w_i$ solves Eq.(\ref{6cone}), so is $uw_i$ for any complex $u$. 
The base of the cone is a manifold $T$
given by the intersection of the space of solutions of Eq.(\ref{6cone}) with a sphere of radius $r$ in ${\cal{\bf{C^4}}}=R^8$, $$\sum_i |w_i|^2 = r^2$$
Treating the 4 $w_i$ as a 4-vector ${\bf w}={\bf x} + i{\bf y}$ on ${\cal{\bf{C^4}}}$ with  2 real vectors ${\bf x}$ and ${\bf y}$, then the above 2 equations yield
\ba
{\bf x} \cdot {\bf x}=r^2/2,  \quad {\bf y} \cdot {\bf y}=r^2/2, \quad {\bf x} \cdot {\bf y}=0
\ea
The first of these equations defines an $S^3$ with radius $r/\sqrt{2}$, while the other 2 equations define an $S^2$ fiber over the $S^3$. Since all such bundles over $S^3$ are trivial, $T$ has the topology of $S^2 \times S^3$. 

We are interested in Ricci-flat metrics on the cone. 
Consider the $ds_5^2$ in the metric (\ref{metric5}), 
$$ ds_5^2 = h_{ab} dx^a dx^b$$
is a metric on $T$, the $6$-dimensional cone admits a Ricci-flat metric if and only if its base manifold admits an Einstein metric, that is,
$R_{ab}(h) = 4 h_{ab}$.
Now let us consider a specific set of $ds_5^2$ metric:
\ba
ds_{T^{P,Q}}^2= {A_0}(\,d\psi + P \cos \theta_1\, d \phi_1 +  Q \cos \theta_2\, d \phi_2   )^2 
+ \sum_{i=1}^2 A_i (\, d\theta_i^2 + \sin^2 \theta_i\, d\phi_i^2\,) 
\ea
where the integers $P,Q$ are coprime. These are metrics on manifolds $T^{P,Q}$ which are fiber bundles over $S^2 \times S^2$ with $U(1)$ fibers. Now, it turns out that only $T^{1,0}$ and $T^{1,1}$ are $S^2 \times S^3$. To be an Einstein metric, $T^{1,1}$ requires $A_0=1/9$ and  $A_1=A_2=1/6$. Furthermore
only $T^{1,1}$ satisfies the Sasaki condition to yield supersymmetry. (A Sasaki-Einstein 5-manifold
may be defined as an Einstein manifold whose metric cone is Ricci-flat and Kahler.) This leads us to study the Sasaki-Einstein manifold $T^{1,1}$ in more detail.

Let us start with the conifold metric with the base manifold $T^{1,1}$, 
\ba
\label{6t11}
ds_{6}^{2}&=& dr^{2} + r^{2} ds_{T^{1,1}}^2  \\
ds_{T^{1,1}}^2 &=&\frac{1}{9}(\,d\psi +\sum_{i=1}^{2} 
\cos \theta_i\, d \phi_i\, )^2 + \frac{1}{6}\sum_{i=1}^{2}
 (\, d\theta_i^2 +  \sin^2 \theta_i\, d\phi_i^2\,) \nonumber  
\ea
It can be shown that $$T^{1,1}=(SU(2)\times SU(2))/U(1)=S^3 \times S^3/U(1)$$ 
which has topology of $S^2 \times S^3$ (with $S^2$ fibered over $S^3$). 
If $\varphi_1$ and $\varphi_2$ are the two Euler angles of the two $S^{3}$s, 
respectively, then their difference corresponds to $U(1)$ while $\psi=  \varphi_1+ \varphi_2$. 
Since $2 \pi \ge \varphi_i \ge 0$, the range of $\psi$ is $ 4 \pi$.

\subsection{The Warped Deformed Conifold}

\begin{figure}[t]
\vspace{1cm}
   \centering
   \hspace{-1cm}\includegraphics[width=2.2in]{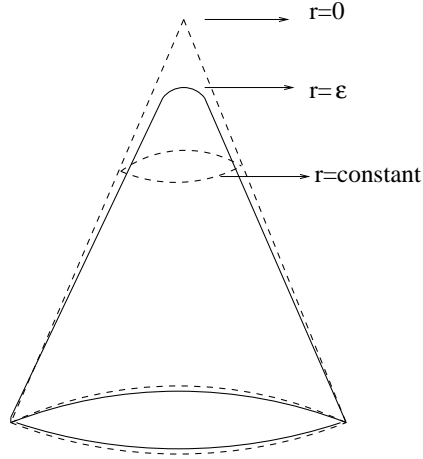}
   \hspace{1cm} 
\caption{Here is a schematic picture of the conifold (dashed line) and the deformed 
conifold (solid line). The apex is at $r=0$. The conifold is deformed at the tip
such that $r=\e$ is now an $S^3$, where $S^{2}$ has shrinked to zero. The dashed circle at constant $r$ represents
the base of the conifold which is a $T^{1,1}$. 
For large $r$, the base of the deformed 
conifold asymptotically approaches $T^{1,1}$.} 
\label{cone}
\end{figure}

The Klebanov-Strassler throat that we are interested in is actually a warped deformed conifold, 
as illustrated in Figure \ref{cone}. This warped deformed conifold emerges in the presence of fluxes.
The R-R flux $F_3$ wraps the $S^{3}$ while NS-NS flux $H_3$ wraps the dual 3-cycle $B$ that generates the warped throat, with warp factor $h(r)$. 
\ba
\frac{1}{ 4\pi^{2} l_{s}^{2}}\int_B H_3 &=& -K, \quad \quad 
\frac{1}{ 4\pi^{2} l_{s}^{2}}\int_{S^{3}} F_3 = M  \nonumber \\
 N &=& KM,  \quad \quad h_{A}=e^{-2\pi K/3g_{s}M}
\ea

Geometrically, the conical singularity of Eq.(\ref{6cone}) can be removed by replacing the 
apex by an $S^3$ \cite{Candelas:1989js},
\ba
\sum_{i=1}^{4} w_i^2=\epsilon^2
\ea
where we shall take $\e$ to be real and small. The resulting deformed conifold is illustrated in 
Figure \ref{cone} and the corresponding metric is non-trivial.
It will be convenient to work in a diagonal basis of the metric, given by
the following basis of 1-forms \cite{Minasian:1999tt,Herzog:2001xk},
\ba
g^1\equiv\frac{e^1-e^3}{\sqrt{2}}~~~~~,~~~~~
g^2\equiv\frac{e^2-e^4}{\sqrt{2}}\nonumber\\
g^3=\frac{e^1+e^3}{\sqrt{2}}~~~~~,~~~~~
g^4\equiv\frac{e^2+e^4}{\sqrt{2}}\nonumber\\
g^5\equiv e^5
\ea
where
\ba
\label{angular}
e^1&\equiv& -\sin \theta_1\, d\phi_1 ~~~~~,~~~~~ e^2\equiv d\theta_1 \, ,
 \nonumber\\
e^3&\equiv& \cos \psi \sin \theta_2\, d \phi_2 -\sin \psi\, d \theta_2\, , 
\nonumber\\
e^4&\equiv& \sin \psi \sin \theta_2\, d \phi_2 +\cos \psi\, d \theta_2\, ,
\nonumber\\
e^5&\equiv& d \psi +\cos \theta_1\, d\phi_1 + \cos \theta_2\, d \phi_2
\ea
The metric of the deformed conifold is studied in 
\cite{Minasian:1999tt,Ohta:1999we,Herzog:2001xk}
\ba
\label{deformedmetric}
ds_6 ^2 = \frac{1}{2}\e^{4/3} K(\tau)\left[\frac{1}{3K^3(\tau)}
(d\tau^2+ (g^5)^2)+\cosh^2(\frac{\tau}{2})[(g^3)^2+(g^4)^2]\right.\nonumber\\
+\left. \sinh^2(\frac{\tau}{2})[(g^1)^2+(g^2)^2]\right]
\ea
where
\ba
\label{K(tau)}
K(\tau)=\frac{(\, \sinh(2\tau)-2\tau\,)^{1/3}}{2^{1/3}\sinh \tau} \, 
\ea
At $\tau \to 0$, the $S^{2}$ (the $g^1$ and $g^2$ components of the metric) shrinks to zero,
and we are left with 
\ba
d\Omega_3^2=\frac{1}{2}\e^{4/3}(2/3)^{1/3}\, \left(\frac{1}{2}(g^5)^2
+(g^3)^2+(g^4)^2 \right)
\ea
which is a spherical $S^3$.
Turning on fluxes, the ten-dimensional metric takes the following ``warped'' form
\ba
\label{10Dwarp}
ds_{10}^2 = h^{-1/2}(\tau) \eta_{\mu \nu}\,dx^{\mu} dx^{\nu}+
h^{1/2}(\tau)\, ds_6^2
\ea
where $ds_6^2$ is given above in Eq.(\ref{deformedmetric}). 
The warp factor $h(\tau)$ is given by the following integral expression \cite{Klebanov:2000hb}
\ba
\label{h(tau)}
h(\tau)= 2^{2/3}\,(g_s M \alpha')^2\, \e^{-8/3}\, I(\tau)\, ,
\ea
\ba
\label{I(tau)}
I(\tau)\equiv\int_{\tau}^{\infty} d\, x \frac{x\coth\, x -1}{\sinh^2 \,x }
\left(\,\sinh(2x)-2x\, \right)^{1/3} \, .
\ea
The asymptotic behavior of $K(\tau)$ and $h(\tau)$, which will be important in
the following calculations, are
\ba
\label{asym}
K(\tau \rightarrow 0)\rightarrow \, (2/3)^{1/3}+{\cal{O}}(\tau^2), \quad
K(\tau \rightarrow \infty)\rightarrow \, 2^{1/3}\, e^{-\tau/3}\nonumber\\
I(\tau \rightarrow 0)\rightarrow \, a_0 +{\cal{O}}(\tau^2), \quad
I(\tau \rightarrow \infty)\rightarrow \, 3\, .\, 2^{-1/3} 
\left(\tau-\frac{1}{4} \right) e^{-4\tau/3}
\ea
where $a_0 \sim 0.71805$. For large $\tau$,
its relation to $r$ in  $ds_6^2$ (\ref{metric5}) is given by
\ba
\label{r}
r^{2}=\frac{3}{2^{5/3}}\e^{4/3}e^{2\tau/3}
\ea
In the limit $\e \to 0$, finite $r$ implies $\tau \to \infty$, so
\ba
ds_{6}^{2} \to dr^{2} + r^{2} \left(\frac{1}{9} (g^{5})^{2} +\frac{1}{6} \sum_{i=1}^{4}  (g^{i})^{2} \right) 
\ea
which is simply the conifold metric (\ref{6t11}).

In a realistic model, the throat does not extend indefinitely. At some point it should glue smoothly to the bulk of the Calabi-Yau manifold. The bulk of the Calabi-Yau manifold is defined to have no significant warping, such 
that the warp factor is essentially constant. Suppose $\tau_c$ is the point where the throat is glued to the bulk. From Eqs.(\ref{h(tau)}), (\ref{I(tau)})
and the asymptotic form of the warp factor for large $\tau$ in Eq.(\ref{asym}), we can find an expression for $\tau_c$ as the size of the throat. Demanding $h(\tau_c)\sim 1$ gives
\ba
\label{tauc}
\exp(\tau_c/3) \sim \sqrt{g_s\, M}\, l_s \e^{-2/3}  \, .
\ea
This relation between $\tau_c, \e$ and $l_s$ is important to understand the spectrum of the graviton
in the next sections.

The physical size of the throat is given by
\ba
\label{throatl}
l=6^{-1/2}\, \e^{2/3}\, \int_0^{\tau_c} d\tau \, \frac{h^{1/4}(\tau)}{K(\tau)} \, 
\sim \sqrt{g_s M}\, l_s \, \tau_c
\ea

We assume that $l \ll V_6^{1/6}$, where $V_6^{1/6}$ is the scale of the compactification. This can be
achieved by a finite value of $\tau_c$ and $l_s \ll V_6^{1/6}$.  Note that the warp factor notations
are related by $e^{2A}= h^{-1/2}(\tau)$.

\section{KK Spectrum of Graviton in a Warped Deformed Conifold}

As explained in Section 2, the graviton KK spectrum, Eq.(\ref{eom1}) or equivalently Eq.(\ref{eom2}), can be obtained from the following action for a scalar field in ten dimensions
\ba
\label{scalaraction}
S=\int d\, x^{4} d\, y^{6} \sqrt{g}(\partial_{M}\Psi_{({\bf m})}
\partial^{M}\Psi_{({\bf m)}}-\frac{1}{2} {\mu}^2\Psi_{({\bf m)}}^2)
\ea
 where for graviton $h_{\mu\nu}=\sum_{{\bf m}} h^{({\bf m})}_{\mu\nu} \Psi_{({\bf m})}$ and ${\mu}=0$. However, to study the spectrum of massive closed strings, we also added the mass term ${\mu}$ which contains the contributions of the closed string oscillatory modes to the mass
\ba
\label{oscmass}
{{\mu}}^2 = 4( {\cal N}-1)/l_s^2 \, .
\ea
For graviton KK modes, ${\cal N}=1$  and ${\mu}=0$, as expected.  
The equation of motion is
\ba
\label{eom}
\frac{1}{\sqrt{g}}
\partial _{M}\left(\sqrt{g}\, \partial ^{M}\, \Pm \right)
-{\mu}^2\, \Pm=0\, .
\ea

We are interested in calculating the KK mass spectrum in the KS throat glued to the bulk
of Calabi-Yau manifold.  To keep the discussion general we consider the metric (\ref{10Dwarp}).
Using this metric, Eq.(\ref{eom}) is transformed to
\ba
\label{wave2}
 \frac{1}{\sqrt{g}} \partial _{y^i}\left(\sqrt{g}\, \partial ^{y^i}\, \Psi_{(m)} \right)
 +({E_{\bf m}}^2 h^{1/2}-{\mu}^2)\, \Psi_{(m)}=0 \quad \quad i=1, ...,6
\ea
where $E_{\bf m}$ is the mass
as  measured by the four-dimensional observer: 
$\eta^{\mu \nu}\partial_{\mu}\partial_{\nu}\Pm=E_{{\bf m}}^2 \Pm$.

Let us cast Eq.(\ref{wave2}) into a standard Schr\"{o}dinger equation 
form that will be used to determine the energy spectrum. Upon expanding the
Laplace operator in Eq.(\ref{wave2}) we will get $\Psi''$ and $\Psi'$, where
the derivative is with respect to the coordinate $\tau$. To transform 
Eq.(\ref{wave2}) into the form of an ordinary Schr\"{o}dinger equation, we perform a 
field redefinition like what usually is done in the hydrogen atom.
Let us decompose the six internal coordinate $y^i, i=1,...,6$ into the
``radial'' coordinate $\tau$ and five azimuthal coordinate $\theta^a, a=1,...,5$ of the base manifold
(like the angular coordinates of $T^{1,1}$ in Eq.(\ref{angular}))
\ba
\label{decompose1}
\Pm \equiv A(\tau) 
B_{({\bf m})}(\tau)\, \Phi_{({\bf m)}}(\theta^a)
\ea
The final wave equation in the radial coordinate $\tau$ will be for function 
$B(\tau)$, while $A(\tau)$ is inserted to get rid of the first derivative
$\Psi'$ term in the wave equation, as explained above. 
The metric in the KS throat takes the following factorized form
\ba
\label{K_mn}
g_{ab}(\tau,\theta)\equiv K_{ab}(\tau) \, \gamma_{ab}(\theta) \, .
\ea
where $\gamma_{ab}$ represents the intrinsic metric component of
the azimuthal directions with only $\theta$ dependence.
Inserting Eq.(\ref{decompose1},\ref{K_mn})
into Eq.(\ref{wave2}), we find
\ba
\label{Schr1}
B_{({\bf m})}''(\tau)-V_{eff}({\bf m},E_{\bf m},\tau)\, B_{({\bf m})}(\tau)=0
\ea
where
\ba
\label{Veff1}
V_{eff}({{\bf m}},E_{\bf m},\tau) =
-g_{\tau\tau}\left({E_{\bf m}}^2 h^{1/2}-{\mu}^2 +\sum_{a,b=1}^{5}\,
K^{ab}(\tau)\,
\frac{{{\mathcal{O}}^{ab}}\Phi_{\bf m}}{\Phi_{\bf m}}\right) +\frac{(G^{1/2})''}{G^{1/2}}
\ea
and $G(\tau)$ up to an overall numerical factor is given by 
\ba
\label{Gtau}
G(\tau)\equiv \sqrt{g}g^{\tau\tau} \, .
\ea
The remaining part of the wavefunction is 
\ba
A(\tau)=G(\tau)^{-1/2} \, .
\ea
The operator ${{\mathcal{O}}^{ab}}$ is defined by
\ba
\label{Ooperator}
\frac{1}{\sqrt{g_{5}}}\partial\, _{\theta^a}
\left(\sqrt{g_{5}}\, \partial\, ^{\theta^a}\, \Phi \right)
&\equiv&
\sum_{m,n=1}^{5}\frac{K^{ab}(\tau)}{\sqrt{\gamma_{5}}}\partial\, _{\theta^a}
\left(\sqrt{\gamma_{5}}\, \gamma^{mn}\partial\, _{\theta^b}\, \Phi \right)\, ,
\nonumber\\
&\equiv& 
\sum_{a,b=1}^{5}\,K^{ab}(\tau)\,{{\mathcal{O}}^{ab}}\Phi\, .
\ea
where ($\gamma_5$) $g_5$ is the (intrinsic) determinant of the azimuthal directions 
of the throat.
The operator ${{\mathcal{O}}^{ab}}$ is closely related to the Laplace operator
for the azimuthal directions. 
Later on, we shall give ${{\mathcal{O}}^{ab}}$ more explicitly.
Eq.(\ref{Schr1}) is the analog of the usual Schr\"{o}dinger equation 
with the ``energy eigenvalue'' ${\cal{E}}=0$, while the energy $E_{\bf m}$ is a parameter in the effective potential $V_{eff}({\bf n},E_{\bf m},\tau)$. The Schr\"{o}dinger equation will have ${\cal{E}}=0$ only for a discrete set of values of $E_{\bf m}$. This results in the quantization of the energy $E_{\bf m}$. 
Our aim is to solve Eq.(\ref{Schr1})
for a specific geometry of a deformed conifold to find the KK spectrum, 
i.e., the discrete values of $E_{\bf m}$. For some earlier works look at \cite{Krasnitz:2000ir}.

We begin with  the ``S-orbital'' solutions of the 10-D graviton, with $ {\mu}=j-1=l=0$, 
In term of the ${{\mathcal{O}}^{ab}}$ operator this corresponds to 
\ba
{{\mathcal{O}}^{ab}}\Phi=0 \, .
\ea
The effective potential is
\ba
\label{VeffS}
V_{eff}= - \frac{1}{6}\, \e^{4/3}\, \Em^2\, \frac{h(\tau)}{K(\tau)^2} 
+\frac{(G^{1/2})''}{G^{1/2}}\, ,
\ea
and $G(\tau)$, up to an irrelevant normalization, is
\ba
\label{G_tau}
G(\tau)=(\sinh(2\tau)-2\tau\,)^{2/3} \, .
\ea
The total wavefunction is given by
\ba
\label{totalwave}
\Psi_{(\bf m)}(\tau)=G(\tau)^{-1/2}\,B_{(\bf m)}(\tau) \, .
\ea
As explained in Section 2, there always exists
a zero mode, the 4-D graviton, with $\Psi_{g}(\tau)=$ constant. This solution corresponds to 
$B_{g}(\tau)=G(\tau)^{1/2}$ and Eq.(\ref{Schr1}) is trivially satisfied.

\section{Infinite Throat Limit}

For the excited graviton modes  the analytical solution of Eq.(\ref{Schr1}) with the above effective 
potential is not known. To get an idea how the spectrum may look like, we consider the limit that the throat is extended to infinity, i.e. $\tau_c \rightarrow \infty$. In this limit we can solve the Eq.(\ref{Schr1}) numerically.

The spectrum for the first five level are given in the first column of table {\bf 1}. The following empirical
relation is obtained for the spectrum which is true for all states, with an accuracy better than 0.1 percent
\ba
\label{linear}
 {\hat E}_{\bf m}=(1.04\, n + 1.01\,) \quad \quad n=1,2...
\ea
where $ {\hat E}_{\bf m}  \equiv \Em g_sM l_s^2\e^{-2/3}$.
This indicates that the spectrum are equally spaced, a property of the the roots of the Bessel functions.
Furthermore, the spectrum is quantized in the unit of $\e^{2/3}\,/ l_s$, as one expects in the spirit of RS model, where $ \e^{2/3}/l_s \sim h(0)^{-1/4}=h_{A}$ is the warp factor. 

\begin{figure}[t]
\centering
\centerline{\vspace{-0.3cm}\hspace{-13cm}\small{$V_{eff}$}}
   \centering
   \hspace{-1cm}\includegraphics[width=2.5in]{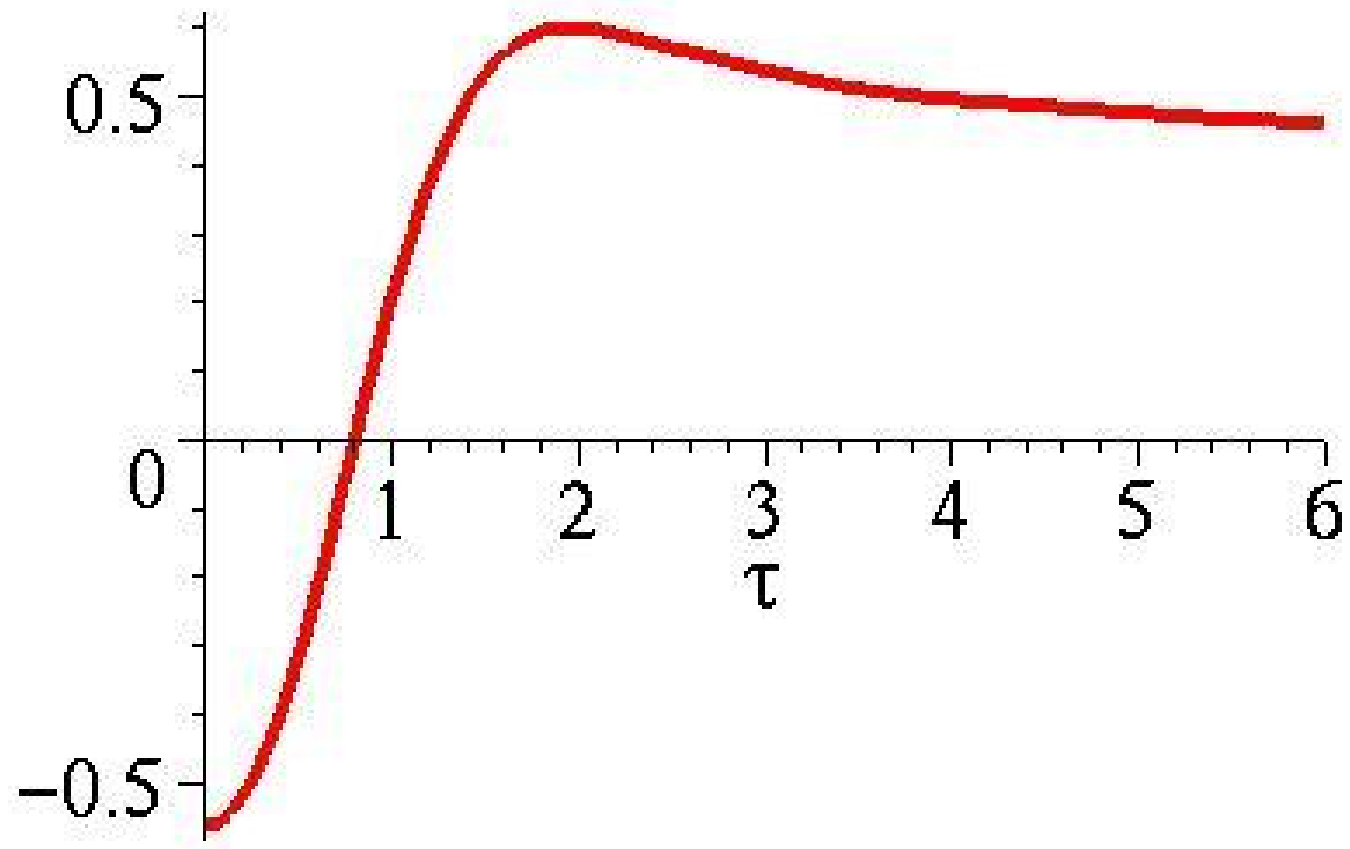}  \hspace{1cm}\includegraphics[width=2.5in]{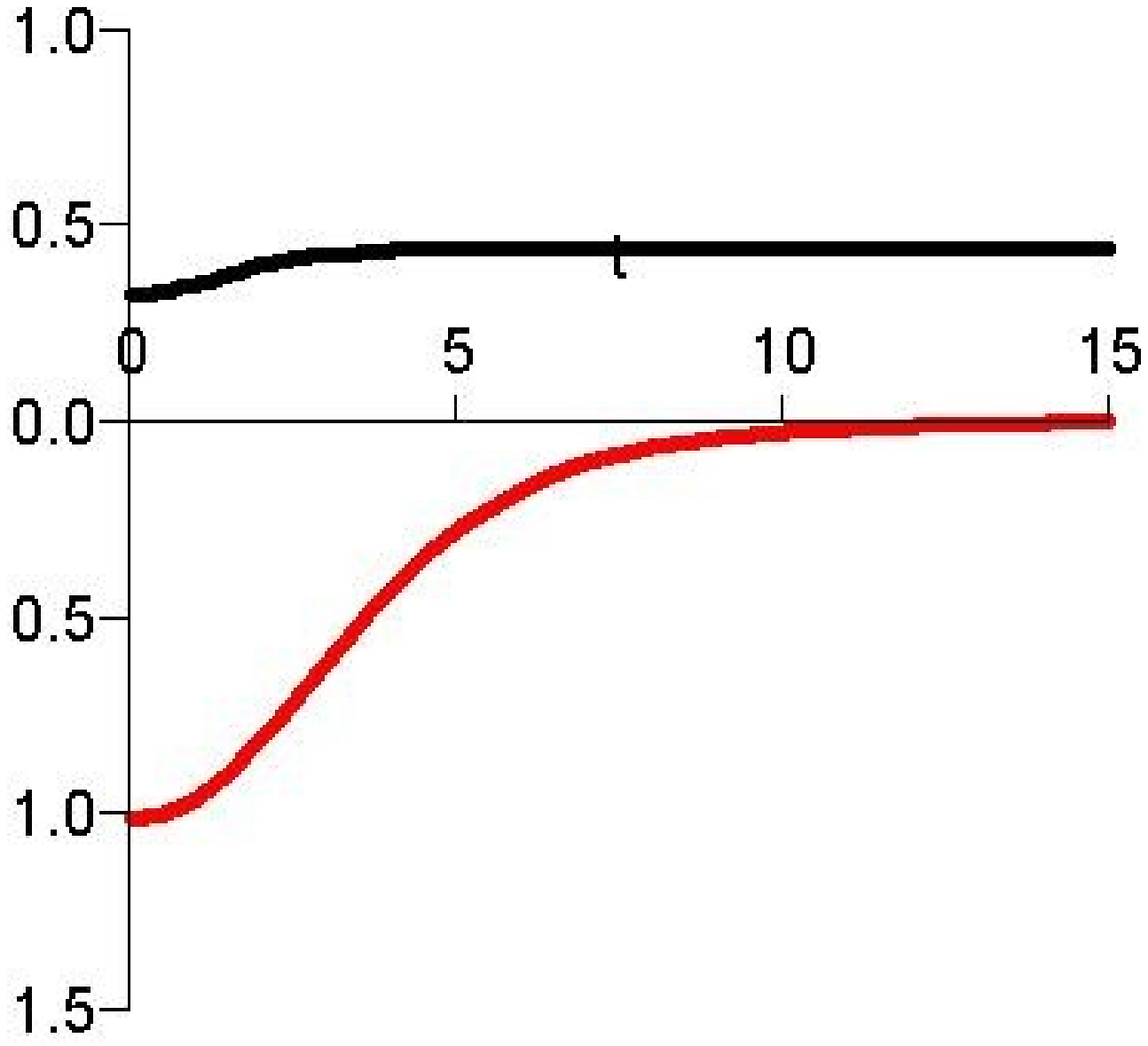}
   \hspace{1cm} 
\centerline{\hspace{-1cm}\small{{\bf{a}}}\hspace{7cm}\small{{\bf{b}}}}
\caption{In the left figure  $V_{eff}(\tau)$ is plotted for $\mu= j-1=l=0$, i.e. excited graviton with no angular momentum. In the right figure, two terms of $V_{eff}$ in Eq.(\ref{VeffS}) for the first excited state are plotted. It is clear that  the last term in $V_{eff}$ quickly reaches the asymptotical value $4/9$, while the first term quickly falls off. This justifies our approximations used in Eq.(\ref{approx1}).}
\label{vplot}
\end{figure}

We can get useful information by looking at the shape of 
the effective potential.  The effective potential is plotted in 
Figure {\bf \ref{vplot}.a}. It has a zero first derivative and a positive
second derivative at $\tau=0$. For large $\tau$ the last term in Eq.(\ref{VeffS}) dominates and 
$V_{eff}$ reaches the asymptotical value $4/9$.
In the Schr\"{o}dinger equation, the effective
energy is zero, which means that the ``particle'' is moving along the 
$\tau$ axis. The energy $\Em$ appears in $V_{eff}$ and only for some quantized
values of $\Em$ the wavefunction falls off at large $\tau$. 

We can perform some reasonable approximations to obtain an analytical solution which captures the above spectrum quite accurately. 
The two components of $V_{eff}$ are plotted in Figure {\bf \ref{vplot}.b }. The function $-I(\tau)/K(\tau)^2$ begins with a constant value $ \simeq 0.94$ at $\tau=0$ and rapidly fall off as $e^{-2\, \tau/3}$ for large $\tau$. The function $(G^{1/2})''/G^{1/2}$ begins with the constant value $2/5$ and quickly approaches its asymptotic value $4/9$.  It is clear that the second term in $V_{eff}$ dominates for large $\tau$ and the solution for $\Bm$ falls off like $e^{-2\tau/3}$.

\vspace{1cm}

\centerline{
\begin{tabular}{|l|c|c|c|r|}  \hline
level     $n$ &     conifold  &roots of $J_{\nu}$	& roots of $J_{2}$&$S^2\times S^3$ \\ \hline
\hspace{0cm} 1& $ 2.05$  & 1.88&1.71	&2.02   \\ \hline
\hspace{0cm} 2&  $ 3.09$ & 2.98&2.81	&3.07   \\ \hline
\hspace{0cm} 3&  $ 4.12$ & 4.05&3.87	&4.11   \\ \hline
\hspace{0cm} 4& $ 5.16$  & 5.11&4.43	&5.15  \\ \hline
\hspace{0cm} 5&  $ 6.20$ & 6.16&5.99	&6.19  \\ \hline
\end{tabular}}

\vspace{1cm}

\noindent {{\bf Table 1.} In this table ${\hat E}_{\bf m}= \Em g_s M l_s^2 \e^{-2/3}$ for the first 5 levels 
of the graviton KK modes with no angular momentum ($\mu=j-1=l=0$) are  given. The first column contains the actual energy spectrum when the metric of the deformed conifold  Eq.(\ref{10Dwarp}) is used.
In the second column, the approximation Eq.(\ref{approx1}) is used and $\Em$ is determined by roots
of Bessel function $J_\nu$, as in Eq.(\ref{Jroot1}). 
In the third column, the energy is calculated from roots of $J_2$. This gives a good order of magnitude estimate.
In the fourth column, the $S^2\times S^3$ approximation of the deformed
conifold, Eq.(\ref{S2S3}), is used. It is clear that this approximation is very good. In each case, the approximation improves as $n$ increases. }

\vspace{1cm}

We introduce the following approximation
\ba
\label{approx1}
\frac{\e^{4/3} \Em^2}{6}\, \frac{h(\tau)}{K(\tau)^2}\simeq \frac{\lambda_1^2}{9} \, 
e^{-4\tau/3\nu}\quad&&\quad 
 \quad  \frac{(G^{1/2})''}{G^{1/2}} \simeq \frac{4}{9} \quad \quad  \quad    
\ea
where $\lambda_1=a\, g_s M \Em l_s^2 \e^{-2/3}= a {\hat E}_{\bf m}$ and the numerical convention in the first term are chosen such that the final expression takes a simple form.
The parameters $a$ and $\nu$ are undetermined yet. They will be determined by the best fit to the 
spectrum (\ref{linear}).

In this approximation, the differential equation (\ref{Schr1})  can be solved analytically. 
The wavefunction, including the  pre-factor $G(\tau)^{-1/2}$, is
\ba
\label{sol1}
\Pm =\frac{1}{(\sinh(2\tau)-2\tau)^{1/3}}\left[ C\,J_\nu \left(\frac{\lambda_1\, \nu}{ 2}\, e^{-2\tau/3\nu}\right)
+D\,Y_\nu \left(\frac{\lambda_1\, \nu}{ 2}\, e^{-2\tau/3\nu}\right)  \right] \, .
\ea
To have a localized solution for $\tau \rightarrow \infty$, we obtain $D=0$ (since 
$Y_{\nu}(z) \sim z^{-\nu}$ as $z \rightarrow 0$). On the other hand, the denominator factor in $\Pm$ vanishes linearly at $\tau=0$, corresponding to the  fact that the size of $S^2$ vanishes at $\tau=0$.  In order to have a regular solution at the bottom of the throat, we obtain, at $\tau=0$,
\ba
\label{rootJ}
J_\nu(\frac{\lambda_1\, \nu}{ 2}) =0 \, .
\ea
This indicates that the spectrum is quantized by the roots of the Bessel function. The roots $x_{\nu\,n}$ 
of $J_\nu$ satisfy the following linear relation for large $n$
\ba
\label{Jroot1}
x_{\nu \, n} \sim n\, \pi +(\nu -\frac{1}{2})\, \frac{\pi}{2} \, .
\ea
To match with our empirical result Eq.(\ref{linear}), this determines
\ba
\label{nua}
\nu \sim 2.44 \quad \quad \quad \quad a \sim 2.48 \, .
\ea
This approximate spectrum from Eqs.(\ref{rootJ}), (\ref{Jroot1}), (\ref{nua}) is given in the second column of table {\bf 1}. The accuracy is very good, and rapidly improves when $n$ increase. The reason that
$\nu\neq 2$ is that the geometry is not pure Ads in the throat. 

This indeed shows that the spectrum is quantized in the unit of $\e^{2/3}/(g_s M l_s^2)$ like in the RS scenario. In other words, compared to the fundamental scale  $l_s^{-1}$, the physical mass is red-shifted by the warp factor $\e^{2/3}/(g_s M l_s)  \sim h(0)^{-1/4}=h_A$, as expected.

Having obtained the spectrum of the graviton with no angular momentum in the 
azimuthal directions, we like to consider their effects to the graviton KK spectrum.
One of the technical issues is the geometrical interpretation
of the azimuthal directions of $ds_6^2$ in Eq.(\ref{deformedmetric}). 
As mentioned before, at $\tau=0$, $g^5, g^4$ and $g^3$ combine to
form a round $S^3$, while $g^1$ and $g^2$ form a configuration
which is topologically a $S^2$ but shrinks to zero size at $\tau=0$. 
For large $\tau$, on the other hand, these five angular coordinates 
asymptotically matches to a $T^{1,1}$. 
To simplify the calculation, we will approximate this complicated 
situation to an easier case when the angular part is $S^2\times S^3$.
We expect this to be a reasonable approximation.
The line element of the metric has the following approximate form
\ba
\label{S2S3}
ds^2= h^{-\frac{1}{2}} \eta_{\mu \nu}\,dx^{\mu} dx^{\nu}+
\frac{\e^{4/3}}{2}\,h^{\frac{1}{2}} K(\tau)\left[\frac{d\tau^2}{3K^3(\tau)}
+\cosh^2(\frac{\tau}{2})\,d\Omega_3^2
+ \sinh^2(\frac{\tau}{2})\,d\Omega_2^2\right]\nonumber\\
\ea
where now $d\Omega_2^2$ and $d\Omega_3^2$ represents the line elements on unit
$S^2$ and $S^3$, respectively.

We can expand the angular dependent part of wavefunction $\Phi_{ ( {\bf m})} $ in 
Eq.(\ref{decompose1}) in terms of the spherical harmonics of $S^2$ and $S^3$,
$Y^{(l\,m)}$ and $Q^{(j\,l'\,m')}$, respectively.
(See the Appendix for a review of the $S^{3}$ harmonics.)
\ba
\label{decompose2}
\Phi_{ ( {\bf m} )}=Y^{(l\,m)}\, Q^{(j\,l'\,m')}
\ea
The operator ${{\mathcal{O}}^{a b}}$, defined in Eq.(\ref{Ooperator}), 
now decompose to products of Laplace
operators for $S^2$ and $S^3$
\ba
K^{a b}(\tau){{{\mathcal{O}}^{a b}}\Phi_{( { \bf m} )}}=
-\frac{2}{\e^{4/3}h^{1/2}(\tau)\,K(\tau)}\left(
\frac{j^2-1}{\cosh^2(\frac{\tau}{2})}+
\frac{l(l+1)}{\sinh^2(\frac{\tau}{2})}\right) {\Phi_{( { \bf m}) }}
\ea

It is easy to add the mass term ${\mu}$ in the potential. As explained before, this 
corresponds for higher excited states with oscillation numbers ${\cal N} > 1$.
The effective potential from Eq.(\ref{Veff1}) becomes
\ba
\label{Vefftotal}
V_{eff}=
  \frac{\e^{4/3}}{6}\,  \frac{h(\tau)}{K(\tau)^2}(-\Em + {\mu}^2\, h^{-1/2}(\tau) )
+\frac{j^2-1}{3\,K(\tau)^3\,\cosh^2(\frac{\tau}{2})} \nonumber \\
+\frac{l(l+1)}{3\,K(\tau)^3\,\sinh^2(\frac{\tau}{2})}
+\frac{(G^{1/2})''}{G^{1/2}}\,\nonumber\\
\ea
where now 
\ba
G(\tau)=(\sinh(2\tau)-2\tau)^{7/6} 
\cosh(\frac{\tau}{2})\sinh(\tau)^{-3/2} \, .
\ea
In the limit that $\cosh(\frac{\tau}{2})K(\tau)^{3/2}=1$ and $j-1=l=0$ the above
expression for effective potential reduces to that of Eq.(\ref{VeffS}).
As a check of the accuracy of our approximation in taking the deformed conifold
to be $S^2\times S^3$, we present the first few energy levels using the above potential
with $j-1=l=0$ and compare them to the previous results in table {\bf 1}.
 
As before, we need to approximate the above effective potential to obtain analytical solutions.
The functions $K(\tau)^{-3}\,\cosh(\frac{\tau}{2})^{-3}$ and $K(\tau)^{-3}\,\sinh(\frac{\tau}{2})^{-3}$
quickly reach the asymptotic value $2$. 
The function ${h^{1/2}(\tau)}/{K(\tau)^{2}}$ varies slightly like $\tau^{1/2}$. So it is a good approximation
to treat these terms as constant in $V_{eff}$ in Eq. (\ref{Vefftotal}).

To obtain the spectrum, we use the WKB approximation
\ba
\label{WKB1}
\int_{\tau_1}^{\tau_2} d\, \tau \sqrt{-V_{eff}}=(n-\frac{1}{2}+
\frac{\delta_{0,l}}{4})\pi \quad \quad n=1,2...
\ea
where $n$ represents the radial quantization of the energy $\Em$ spectrum in the warped throat.
When $l=0$, $\tau_1=0$ and $\tau_2$ is the point where $V_{eff}(\tau_2)=0$. 
The factor $-1/4$ comes from the fact that the wave function vanishes at 
the center, $\tau=0$. When $l\neq 0$, the equation $V_{eff}=0$ has 
two roots : $\tau_1$ and $\tau_2$. 

Performing the integral (\ref{WKB1}), we obtain
\ba
\label{WKBE} 
\sqrt{  {\hat E}_{\bf m} ^2-{\cal{Q}}^2}-
{\cal{Q}}\,\arctan \left(\frac{\sqrt{   {\hat E}_{\bf m}^2-{\cal{Q}}^2}}{{\cal{Q}}}\right)
=1.03 (n-\frac{1}{2} + \frac{\delta_{0,l}}{4})
\ea
where 
\ba
\label{Q}
{\cal{Q}}^2\equiv   c\, {\hat{\mu}}^2 +d\, (j^2-1) +f\, l(l+1) + .65\, ,
\ea
and 
\ba
\label{mu}
{\hat{\mu}}^2  \equiv  \frac{1}{2^{2/3}3}\,(g_s\,M\, l_s^2)\, {{\mu}}^2=
\frac{2^{4/3}}{3}\, g_s M {\cal{N}} \, .
\ea

\centerline{
\begin{tabular}{|l|c|c|c|c|r|}  \hline
 $n$& $\hat{\mu}$ & $j-1$& $l$ & ${\hat E}_{\bf m}$  &  WKB          \\ \hline
 1  &  0    &  0   & 0   & 2.02 &  1.87          \\ \hline
  1  &  0    &  1   & 0   & 3.61 &  3.53          \\ \hline
 2  &  0    &  1   & 0   & 4.71 &  4.76          \\ \hline
 1  &  0    &  2   & 0   & 5.09 &  4.93         \\ \hline
 1  &  0    &  0   & 1   & 4.00 &  3.77          \\ \hline
 2  &  0    &  0   & 1   & 5.06 &  5.10          \\ \hline
 1  &  1    &  0   & 0   & 3.38 &  3.31          \\ \hline
  2  &  1    &  0   & 0   & 4.57 &  4.53        \\ \hline
  1  &  1.3  &  0   & 0   & 3.97 &  3.90        \\ \hline
  1 &  1    &  1   & 0   & 4.39 &  4.30           \\ \hline
  1  &  1    &  1   & 1   & 5.81 &  5.02       \\ \hline
  \end{tabular}}

\vspace{1cm}

\noindent {{\bf Table 2.} In this table, a sample of the states in the energy spectrum is given when
various quantum numbers are present. Column labeled by ${\hat E}_{\bf m}$ contains the actual dimensionless value of energy, while the column labeled by WKB contains the estimates from the WKB approximation. In general the accuracy of WKB is around a few percent.

\vspace{1.5cm}

A good fit to the data is given by
\ba
\label{datafit}
c \simeq 3.4 , &\quad d \simeq 1.4 , \quad f \simeq 3.4  \, .
\ea
In table {\bf 2}, the result for ${\hat{E}_m}$ are given. The accuracy of our WKB approximation is in
general quite good.
In the limit that $\mu=j^2-1=l=0$, Eq.(\ref{WKBE}) rapidly approaches the linear regime
Eq.(\ref{linear}). Also, when one of the quantum number $\mu, j $ or $l$ is turned on, Eq.(\ref{WKBE})
also reaches a linear regime between $\Em$ and that quantum number.

As an estimate of $\hat{\mu}$, we use the numerical values applied in KKLMMT, $M=20$, $g_s=0.1$, which give
\ba
\hat{\mu} \sim 1.3\, \sqrt{{\cal N}-1}\, .
\ea
So $\hat{\mu}$ is of order unity for the first few excited oscillatory modes. 

\section{A Finite Throat Glued to the Bulk}

Now that we have obtained the KK gravity spectrum for an infinite warped deformed conifold,
we like to consider the more realistic case when the throat has a finite size and is glued 
to the bulk at $\tau_c$. We are interested in the magnitude of a typical
KK mode wave function outside the throat. Towards the end of brane inflation, the KK modes are produced in the inflation (i.e., the A) throat. Naturally, these KK modes are localized at the bottom of the A throat. Suppose the standard model branes are sitting somewhere else, either as D7-branes wrapping 4-cycles in the bulk, or as a stack of 3-branes sitting in another throat. In either case, we like to estimate the tunneling amplitudes of the KK modes, that is, their amplitudes in the bulk and/or in the standard model thraot. If the tunneling amplitude is very much suppressed, then these KK modes will mostly decay into gravitons in the A throat, leading to an early universe dominated by gravitational radiation, an undesirable situation. In this section, we are interested only in order of magnitude estimates.

The details of the bulk is not important. The compactified manifold can have a number of throats. The key properties of the bulk relevant for our investigation are (1) it is either weakly warped or not warped at all; (b) it has a finite size. Both of these features should be quite generic in a realistic model. 
To simplify the problem, we like to consider a toy model which embodies the above  features.

We can obtain a simple compact manifold by gluing an identical conifold to the original one at $\tau=T$ and then perform an orientifold projection. 
The range of coordinate $\tau$ is $0 \le \tau \le T$. For $0 \le \tau \le \tau_{c}$, we keep the same warped deformed conifold described earlier. Here, at $\tau_{c} \le T$, $h(\tau_{c}) = 1$. For $\tau_{c} \le \tau \le T$, we keep the condition $h(\tau) = 1$, that is, there is no warping. So the region $\tau_c \le \tau \le T$ defines our bulk.   
That is, we take the bulk to be part of the original conifold but with $h(\tau) = 1$. 
For large $\tau > \tau_{c}$, $r$ given by Eq.(\ref{r}) is the natural coordinate and the metric is
\ba
\label{bulk}
ds^2 \sim \eta_{\mu \nu} \, dx^{\mu} dx^{\nu} + dr^2 +r^2 ds^2_{T^{1,1}}
\ea
The physical size of the compactification is given by
\ba
\label{consize}
R \simeq \frac{3^{1/2}}{2^{5/6}}\,  \e^{2/3}\, e^{T/3} \ .
\ea
Note that, for $\tau \ge \tau_c$, only the combination $\e^{2/3}\, e^{\tau/3}$ is meaningful. 
For $\tau_c \le \tau \le T$, the effective potential has the same form as in Eq.(\ref{VeffS}), except that
$h(\tau)=1$. Furthermore, $K(\tau)$ falls off like $e^{-2\tau/3}$.

\begin{figure}[t]
\centering
  \hspace{-.5cm} \includegraphics[width=4.in]{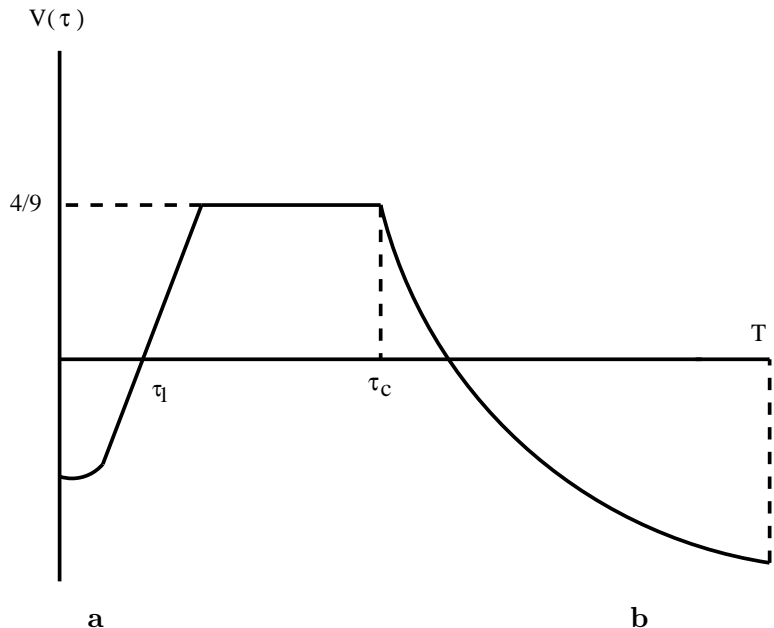}  
  \centerline{\hspace{-1cm}\small{{\bf{a}}}\hspace{7cm}\small{{\bf{b}}}}
\caption{In this figure the qualitative shape of $V_{eff}$ given in Eq.(\ref{Vtotal}) is plotted. 
$V_{eff}$ in the throat region quickly reaches the vale $4/9$. The throat is glued to the bulk at $\tau_c$. 
The bulk ends at $T$, the position
of the O-3 plane. There is a mirror image of this potential for $T < \tau < 2\,T$.} 
\label{pot}
\vspace{1cm}
\end{figure}

To further simplify the exercise, we replace $\nu=2.44$ by its nearest integer, namely, $\nu=2$ for $0\le \tau\le \tau_c$.  This makes the solution more symmetric when the throat is glued to the bulk.  With this choice of $\nu$, the comparison to our empirical result Eq.(\ref{linear}) sets $a=3$. The spectrum for this choice of $\nu=2$ and $a$ is given in the third column of table 1. The approximation gets better for higher $n$ and is sufficient for our purpose here. In addition, we shall consider only S-waves.

The effective potential now becomes
\ba
\label{Vtotal}
V_{eff}=  - \frac{1}{6}\, \e^{4/3}\, \Em^2\, \frac{h(\tau)}{K(\tau)^2} 
+\frac{(G^{1/2})''}{G^{1/2}}\,  \simeq \left\{ \begin{array}{c}
         (-{\lambda_1^2} e^{-2\tau/3} +4)/{9}  \quad\quad 0\le \tau \le \tau_c \\
         ( - {\lambda_2^2} e^{2\tau/3}  +4)/{9} \quad\quad \tau_c \le \tau \le T
         \end{array} \right.      
\ea   
where 
$\lambda_2=3^{1/2} 2^{-5/6} \Em \e^{2/3}$. The shape of the effective potential is plotted in
fig. {\bf {\ref {pot}}}.
The continuity of the potential at $\tau_c$ requires that
\ba
\label{cont}
\lambda_2 = e^{-2\tau_c/3}\, \lambda_1 \, .
\ea
This is equivalent to Eq.(\ref{tauc}), requiring that $h(\tau_c)=1$.
Now, the wave function solution is still the same as in Eq.(\ref{sol1}), but with the new boundary condition at $\tau_c$. The new wave function becomes
\ba
\label{totalwave}
\Pm= \frac{1}{(\sinh(2\tau)-2\tau)^{1/3}}\, \left[ C_{1,2}J_2(\lambda_{1,2} e^{\mp \tau/3}) + 
D_{1,2}Y_2(\lambda_{1,2} e^{\mp \tau/3}) \right]
\ea 
The solution in the throat with $e^{-\tau/3}$ in the arguments of the Bessel functions represent an exponentially falling off solution, while, in the bulk, the Bessel functions with argument  $e^{+\tau/3}$
represent a radially propagating plane wave, which decays only like $r^{-2}$. This can easily be understood by choosing the more natural coordinate $r$ given in Eq.(\ref{r}) for the asymptotical behavior of $J_2$ and $Y_2$.

There are 3 length scales in the model, namely, 
the string scale, $l_s$, the warped scale $\e^{2/3}$, and the size of compactification $R$.
We find useful to introduce 3 dimensionless parameters, $\beta_1$,  $\bm$ and $\beta_3$ :
\ba
\label{beta}
\beta_1 \equiv e^{-\tau_c/3}  \quad , \quad 
\bm  \equiv \lambda_1 e^{-\tau_c/3} \quad , \quad
\lambda_2\,  e^{T/3} \equiv \frac{\bm}{\beta_3}=\beta_{23}
\ea
where we have also introduced the notation $\beta_{ij}=\beta_{i}/\beta_{j}$.
Here, $\beta_1 \sim \e^{2/3}/\sqrt{g_s M} l_s \sim h(0)^{-1/4}=h_{A}$ is simply the warp
factor at the bottom of the throat. $\beta_2 \sim 3\sqrt{g_s M}\, \Em l_s$, which measures the energy 
(the KK mass) in term of the string  scale. Finally $\beta_3 \sim l_s/R$ measures the scale of the compactification in terms of the string scale. Given $\beta_{1}$ and $\beta_{3}$, we like to solve for
$\beta_{2}$, i.e., the KK spectrum.
In order to trust the low energy supergravity approximation, $\beta_3\ll 1$. 
Also, for small warp factor, $\beta_1 \ll1$.
We find that there is a competition between  the
smallness of $\beta_1$ and $\beta_3$. If $\beta_1 \ll \beta_3$, the spectra are determined by
the warped throat; we refer to this as throat domination. On the
other hand, if $\beta_3\ll \beta_1$, the spectra are determined by the overall size of the compactification. This is the large extra dimension limit and we refer to it as the bulk domination.
In the intermediate region both $\beta_1$ and $\beta_3$ play important roles. Finally,
for the situation of interest, we also have $\bm \ll 1$. This is because the energy is reduced
compared to the natural scale $l_s^{-1}$ either by warping in the throat domination case
or by the large volume effect as in the bulk domination case. This approximation for $\bm$ breaks down
when the warping or the scale of compactification do not create a hierarchy of scales. In this discussion, we do not consider very highly excited modes that reach the continuum limit (that is, $\bm \sim 1$). 

Regularity of the solution at $\tau=0$ now requires 
\ba
D_1= -\frac{J_2(\lambda_1)}{Y_2(\lambda_1)} \, C_1 \, = -\frac{J_2(\beta_{21})}{Y_2(\beta_{21})} \, C_1 \, 
\ea
The boundary  condition at $\tau=T$, corresponding to  $r=R$ the position of the O3-plane is given by
Eq.(\ref{bc}). We consider the symmetric solution with respect to  the O3-plane, that is 
$\Pm(\tau)=\Pm(2T-\tau)$. This will set the boundary condition $\Pm'(\tau=T)=0$, which gives
(using the Bessel function recurrence relations)
\ba
D_2 \simeq  -\frac{J_3(\beta_{23})}{Y_3(\beta_{23})} \, C_2 \, .
\ea 

The continuity of the wavefunction and its derivative at $\tau=\tau_c$ will now determine the energy, i.e., $\beta_{2} $,
\ba
\label{C1C2}
C_2=\frac{Y_3(\br) \left[ J_2(\bm) Y_2(\bt)-Y_2(\bm) J_2(\bt) \right ] }
{Y_2(\bt) \left[ J_2(\bm) Y_3(\br)-Y_2(\bm) J_3(\br) \right ] }  \, C_1 \, .
\ea
and
\ba
\label{energy}
\frac{J_2(\bm) Y_2(\bt)-Y_2(\bm) J_2(\bt) }
{J_1(\bm) Y_2(\bt)-Y_1(\bm) J_2(\bt)}
=\frac{J_2(\bm) Y_3(\br)-Y_2(\bm) J_3(\br) }
{J_3(\bm) Y_3(\br)-Y_3(\bm) J_3(\br)} 
\ea

Let us first consider the two limits of Eq.(\ref{energy}). One limit is
the infinite throat studied in the previous section, where the
energy spectrum is quantized by the roots of $J_2$. This corresponds to
$\tau_c \rightarrow T$. In this limit $\beta_3 \rightarrow 1$  and the denominator of right hand side
of Eq.(\ref{energy}) vanishes. This requires that 
\ba
\label{limit1}
J_1(\bm) Y_2(\bt)-Y_1(\bm) J_2(\bt)=0
\ea
For $\bm \ll1$ this gives the expected result
 \ba
 \label{limit1,2}
 J_2(\bt) \simeq 0 \, .
\ea

\begin{figure}[t]
\vspace{1cm}
\centering
  \hspace{-.5cm} \includegraphics[width=3.in]{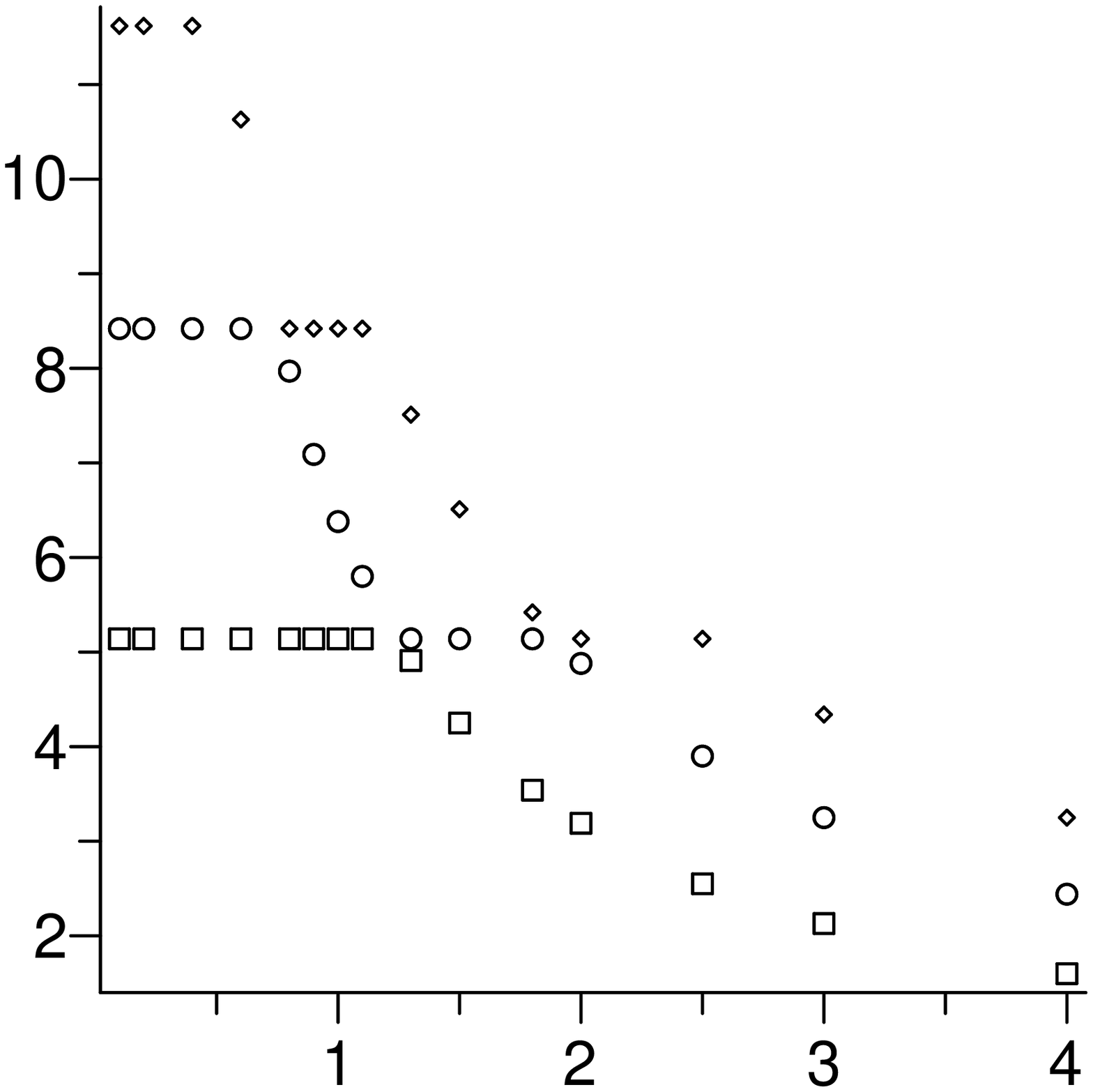}  \includegraphics[width=3.in]{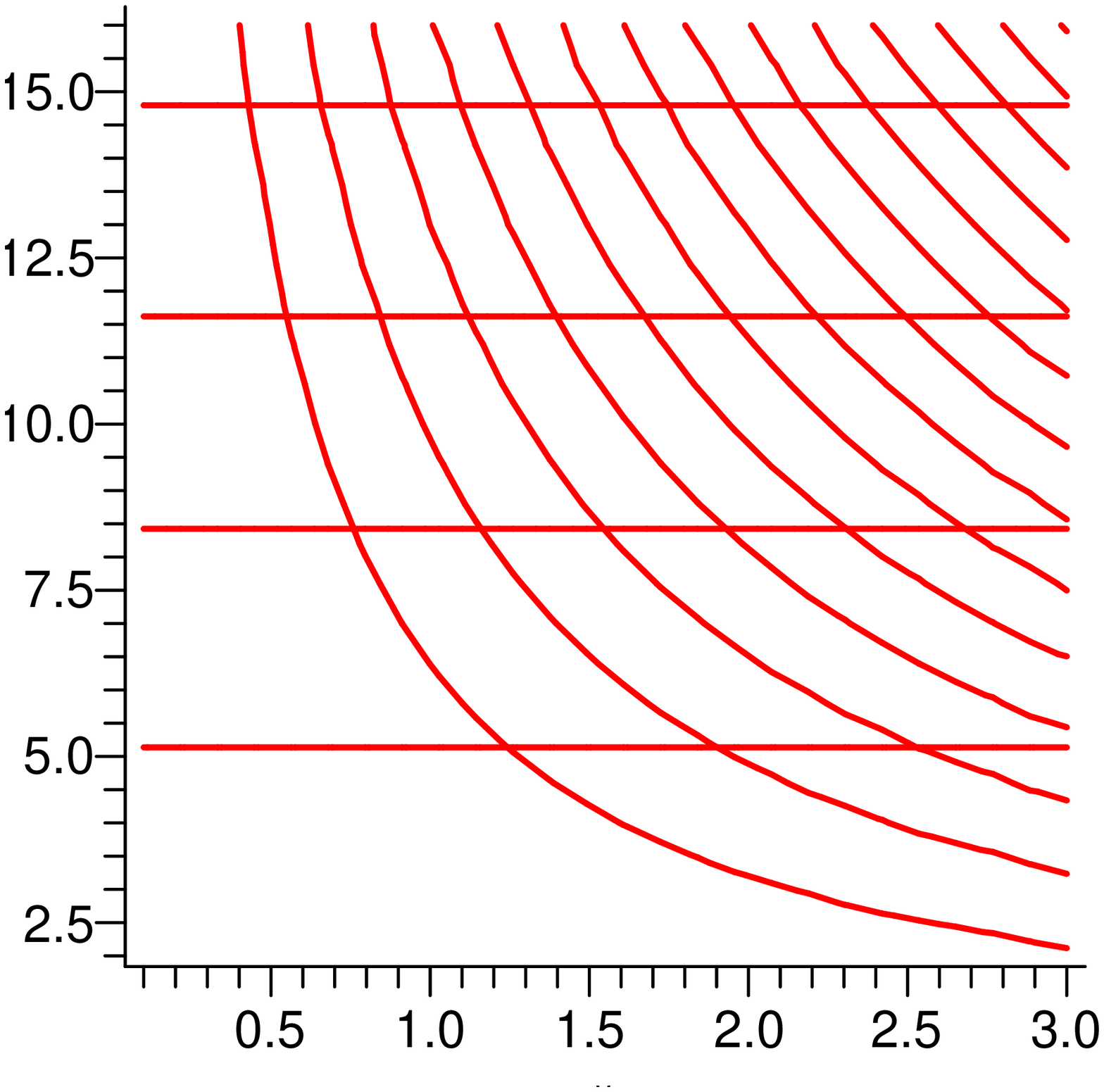}  
  \centerline{\hspace{-1cm}\small{{\bf{a}}}\hspace{7cm}\small{{\bf{b}}}}
\caption{ In Figure {\bf a}, $\beta_{21}$ (measures the KK mass spectrum) versus $\beta_{13}$ is plotted by solving Eq.(\ref{energy}) numerically. The curve formed by box, big circle and small circle represent the first, second and third excited states respectively. 
Here $\beta_1=1/50$ but the shape of the plots do not change if other values
for $\beta_1$ are chosen as input parameter. In the asymptotic region when $\beta_{21} \sim $ constant,  the  throat dominates and the spectrum is given by Eq.(\ref{limit1,2}) (the horizontal lines in {\bf b}) . The other asymptotic region is when the bulk dominates and the spectrum is given by Eq.(\ref{limit2,2})
(the curved lines in {\bf b}). 
In between the spectrum is given by the superposition of these 2 sets of curves.
The transition between these two regions happens earlier for the excited modes.  Also the excited modes halt temporarily when $\beta_{21}$ approaches the following lower roots of $J_2$ and  afterwards drops as given  by Eq.(\ref{limit2,2}). In the right figure the solutions of Eq.(\ref{limit1,2}) and Eq.(\ref{limit2,2}) versus $\beta_{13}$ are plotted. 
The agreement between our numerically obtained curve and the superposition of the theoretical curves is very good. Note that the energies of different KK modes never cross or become degenerate.}
 \label{running}
\end{figure}

The other limit is the no throat case when $\tau_c \rightarrow 0$, corresponding to a deformed
conifold without warping. In this limit, $\beta_1 \rightarrow 1$ and the left hand side of Eq.(\ref{energy})
vanishes. This in turn requires that 
\ba
\label{limit2}
J_2(\bm) Y_3(\br)-Y_2(\bm) J_3(\br)=0 \, ,
\ea
which for $\bm \ll 1$ gives
\ba
\label{limit2,2}
 J_3(\br) \simeq 0 \, ,
\ea
which determines the energy $\bm$ in terms of the compactification $\beta_3$ as expected.
Condition (\ref{limit2}) is a manifestation of the fact that the wavefunction is smooth
at $\tau=0$, when the size of $S^2$ vanishes. To summarize, we see that, in terms of $\beta_{13}$, 
the spectrum is given by Eq.(\ref{limit1,2}) as $\beta_{13} \rightarrow 0$ (the throat domination limit, given by the horizontal line in Fig \ref{running}.b),
and by Eq.(\ref{limit2,2}) as $\beta_{13} \rightarrow \infty$ (the bulk domination limit, given by the horizontal line in Fig \ref{running}.b).
Using Eq.(\ref{Jroot1})  for the $n$th root of $J_2$ and and the $m$th root of $J_3$, we find that
\ba
\label{beta1,3}
\beta_1 \simeq  \left(\frac{4\, m +5}{4\, n +3} \right)\, \beta_3 \,, \quad \quad 
\ea
where $m$ and $n$ represents the first few excited states.

The interesting case is the intermediate region between these two limits.
Now we like to make the following observation:
properly interpreted, the spectrum in the intermediate region is completely determined by 
the combination of Eq.(\ref{limit1,2}) and by Eq.(\ref{limit2,2}). To a very good approximation,
the spectrum for general $\beta_{13}$ is given by the superposition of the 2 sets of curves 
shown in Fig \ref{running}.b.
This is checked numerically. 
In Fig.{\ref{running}.a} we have plotted the first few KK mass $\beta_{21}$ versus
$\beta_{13}$. In the limit that $\beta_1 \ll \beta_3$, the KK mass is quantized by Eq.(\ref{limit1,2}). As $\beta_1$ increases and becomes comparable to $\beta_3$
the curve begins to fall down and for $\beta_1\gg \beta_3$, the KK mass is quantized by Eq.(\ref{limit2,2}). 
The agreement between these two graphs is very good.

\section{Tunneling and Wavefunction Distributions}

\begin{figure}[t]
\centering
  \hspace{-.5cm} \includegraphics[width=5.in]{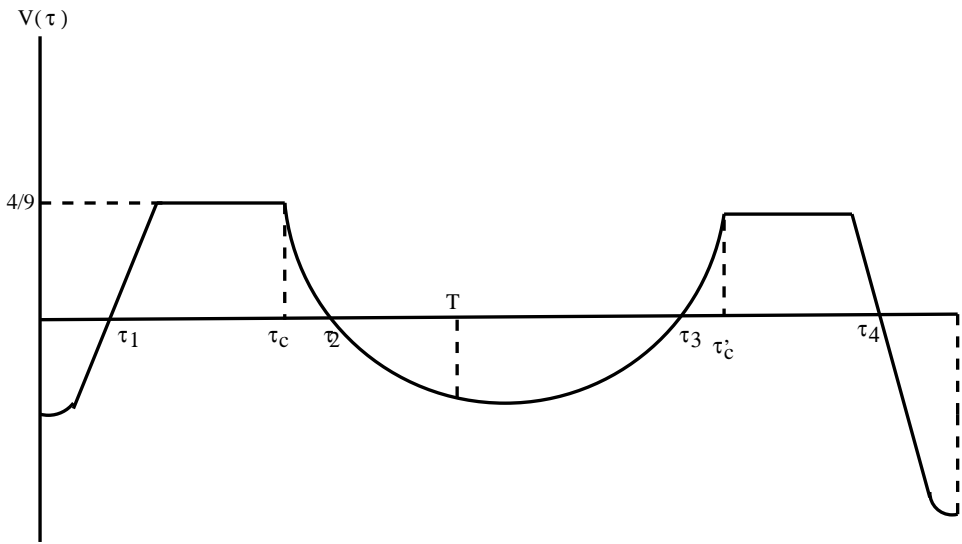}  
\caption{In this figure the qualitative shape of effective potential for the two throats model is shown.
The inflationary A throat begins at $\tau=0$ and ends at $\tau_c$. 
The standard model S throat ends at $\tau'_c$.
In between, $\tau_2< \tau< \tau_3$, is the bulk .}
\label{pot2}
\vspace{1cm}
\end{figure}

In this section we study the two throat model. The inflation takes place in the A throat while the
standard model is located in the S throat. A qualitative view of this two throat model is sketched in
Fig.{\bf{\ref{pot2}}}. The effective potential is given in Eq.(\ref{Vtotal}). Compared to the
set up in the previous section we drop the $Z_2$ symmetry on the throats and the throats now 
have different warp factors.
As before, the energy ${\cal E}$ in our Schr{\"o}dinger equation is zero. There are two barriers
given by region $\tau_1 < \tau <\tau_2$ and $\tau_3 < \tau <\tau_4$ 
that the particle must tunnel through.

The tunneling rate, or transmission coefficient, is straightforward to obtain in the WKB approximation.
Beginning with the first barrier, the A throat to the bulk, the relation between the coefficients of the incoming waves and the outgoing waves is given by \cite{Merzbacher}
\ba
\label{matrix1}
\frac{1}{2}\left( \begin{array}{c}
 \Theta_A  +\frac{1}{ \Theta_A}       \quad i\,(\Theta_A  -\frac{1}{ \Theta_A} )\\
-i\,(\Theta_A  -\frac{1}{ \Theta_A})  \quad  \Theta_A  +\frac{1}{ \Theta_A}
\end{array}  
\right) 
\ea
\ba
\Theta_A &=& 2\, \exp\left (\,\int_{\tau_1 }^{\tau_2} d\tau \sqrt{V_{eff}(\tau)}\, \right ) 
\ea
where $\tau_1$ and $\tau_2$ are the classical  turning point, as shown in Fig. \ref{pot2}. 
Using Eq.(\ref{Vtotal}), we obtain
\ba
\label{Theta1}
\Theta_A &=&2 \exp \left(-2 \sqrt{4-\beta_2^2}\right) 
\left(\frac{2+\sqrt{4-\beta_2^2}}{2-\sqrt{4-\beta_2^2}}\right)^2
\nonumber\\
 &\simeq&  e^{-4}  2^9 \beta_2^{-4} \, .
\ea
In the second line the approximation $\beta_2 \ll 1$ is taken.
The transmission coefficient from the A throat to the bulk is given by
\ba
\label{tran1}
T( \mbox{A} \rightarrow \mbox{bulk}) &=&4 \left( \Theta_A  +\frac{1}{ \Theta_A}   \right)^{-2} \nonumber\\
&\simeq&(\frac{e}{4}\, \beta_2)^8
\ea
This may be very small. In the throat domination region, using Eqs.(\ref{linear}), (\ref{nua}) and (\ref{beta}) we obtain 
 \ba
 T(\mbox{ A} \rightarrow \mbox{bulk}) \sim ((n+1)\, h_A)^8
 \ea
where $h_A$ is the warp factor in A throat and $n$ is the radial KK mode.
   
We are also interested in the probability of tunneling from the A throat to the S throat via the bulk, as shown in Fig.{\bf{\ref{pot2}}}. The matrix relating the 
coefficients of the incoming wave from the A throat to the outgoing wave in the S throat is given by
\ba
\label{matrix2}
\frac{1}{4}\left( \begin{array}{c}
 \Theta_A  +\Theta_A^{-1}       \quad i\,(\Theta_A  - \Theta_A^{-1}  )\\
-i\,(\Theta_A  -\Theta_A^{-1})  \quad  \Theta_A  +\Theta_A^{-1}
\end{array}  
\right)  
\left( \begin{array}{c}
e^{-i\, L} \quad 0\\
0 \quad e^{i\, L} 
\end{array}  
\right)  
\left( \begin{array}{c}
 \Theta_S  +\Theta_S^{-1}       \quad i\,(\Theta_S  -\Theta_S^{-1})\\
-i\,(\Theta_S  -\Theta_S^{-1})  \quad  \Theta_S  +\Theta_S^{-1}
\end{array}  
\right)  \nonumber\\
\ea
where $L$ is the following phase integral over the bulk
\ba
L=\int_{\tau_2}^{\tau_3} \sqrt{-V_{eff}(\tau)}\, d\tau 
\ea
To get some quantitative idea about the value of $L$, we may assume that the bulk is 
symmetric between two throats at $\tau=T$, so
\ba
L&=&2  \int_{\tau_2}^{T} \sqrt{-V_{eff}(\tau)}\, d\tau \nonumber\\
&=&2 \sqrt{\beta_{23}^2 -4} - 4\, \mbox{arctan}(\frac{\sqrt{\beta_{23}^2 -4}}{2}) \, .
\ea 
Also $\Theta_A=\Theta_S$ (independent of $Z_2$ symmetry of the bulk)
by noting that $\Theta_S$ is given by the same expression as in Eq.(\ref{Theta1}),  i.e., $\Theta_S$ is only a function of the KK mass, $\beta_2$, which must be the same in both throats. This means 
$E_{{\bf m}}(A)=E_{{\bf m'}}(S)$, although the KK quantization mode $n_{A}$ is in general different from $n_{S}'$.
For example when both throats are large enough so Eq.(\ref{linear})
holds, we obtain $(n_A+1) h_A = (n_S+1) h_S$, where $h_S$ is the warp factor in the S throat.
So the transmission coefficient from the A throat to the S throat via a bulk is given by
\ba
T (A \rightarrow S) & =& 4\, \left( ( \Theta_{A} \Theta_{S} + \frac{1}{\Theta_{A} \Theta_{S}})^{2} 
\cos^{2} L + (\frac{\Theta_{A}}{\Theta_{S}} +\frac{\Theta_{S}}{\Theta_{A}})^{2} \sin^{2} L\right)^{-1}
\nonumber\\
&=&4\, \left( ( \Theta_{A}^2 + \frac{1}{\Theta_{A} ^2})^{2} 
\cos^{2} L + 4\, \sin^{2} L\right)^{-1}
\ea
In the absence of the bulk, $L=0$ and $T (A \rightarrow S)$ is very small,
\ba
T (A \rightarrow S) \sim (\Theta_A )^{-4} \sim (\frac{e}{4}\, \beta_2)^{16} \sim (n_{A}h_{A})^{16}
\ea
This should be compared to the Randall-Sundrum 5-dimensional spacetime scenario where  
 $T (A \rightarrow S)  \sim (n_{A} h_A)^4$ \cite{Dimopoulos:2001ui}.

However, for some KK modes, $L=(n_{L}+1/2) \pi$, so $\cos L =0$.
For these KK modes, the transmission coefficient approaches unity 
\ba
T (A \rightarrow S)  \sim 1
\ea
This is the well-known resonance effect. For this resonance effect to take full place, 
\ba
\label{res1}
2 \sqrt{\beta_{23}^2 -4} - 4 \mbox{arctan}(\frac{\sqrt{\beta_{23}^2 -4}}{2}) =(n_L+\frac{1}{2})\, \pi
\ea
Suppose we are in the bulk domination scenario and the energy quantization is given by Eq.(\ref{limit2,2}).
For very massive KK mode, $\beta_{23}\gg1$; using Eq.(\ref{Jroot1}) for the Bessel functions, we find
that the resonances always occur. For these modes 
\ba
\label{res2}
n=-1+\frac{n_L}{2}
\ea
This means that tunneling from the A throat to another throat passing through the bulk may not be suppressed at all. In any realistic model, the KK modes would have some widths. Suppose $h_{S} \ll h_{A}$. We expect a $n_{A}$th KK mode will overlap with a small set of $n_{S}$th KK modes so that the resonance effect will come into play.

Another quantity of phenomenological interest 
might be the ratio $P$ of a KK mode $\Pm$ in the bulk to its value at $\tau_1$ the classical turning point
in Fig. {\bf{\ref{pot2}}}:
\ba
\label{P}
P \equiv |\frac{\Pm(\tau_c)}{\Pm(\tau_1)}|^2 = \frac{\beta_2^4}{16} \,
\left(\frac{J_2(\bm) Y_2(\bt)-Y_2(\bm) J_2(\bt)}{J_2(2) Y_2(\bt)-Y_2(2) J_2(\bt)}\right)^2
\ea
In the throat dominated limit, the term in the bracket scales like $\beta_2 ^2$ and $P$ scales like
$\beta_2^8 \simeq ((n+1)\,h_{A})^{8}$ for the  radial excitation $n$. Since the typical warp factor in brane inflation has value
$h_{A} \sim 1/5$ to $10^{-3}$, we see that $P$ can be exponentially small.
However, in the bulk dominated region, the term in the bracket  scales like $\beta_2^{-2}$ and 
$P$ increases rapidly and may become comparable to one. 
We see that there is little suppression due to tunneling.

Another quantity of interest is the ratio of the averaged wavefunction of the excited modes compared to
 the zero mode 
 \ba
 \label{ratio2}
 {\cal {R}}_1 &\equiv&  \left|\frac{\overline{\Pm}}{\Psi_{g} } \right| ^2
  \ea
  where $\overline{\Pm}$ is the averaged wavefunction in the bulk. 
  To find ${\cal  R}_1$, we need to use the normalized $\Psi_{g}$ and $\Pm$ as given in Eq.(\ref{ortho}).
Assuming $T \gg \tau_c \gg1$, one obtains
  \ba
\label{ratio22}
{\cal {R}}_1 \simeq
 \left[1+\left( \frac{J_2(\bm) Y_3(\br)-Y_2(\bm) J_3(\br)}{J_2(\bm) Y_2(\bt)-Y_2(\bm) J_2(\bt)}\right)^2
 \right]^{-1}
\ea
In the bulk dominated region, one finds (using Eq.(\ref{limit2})) that
\ba
{\cal {R}}_1 \sim 1\, .
\ea
This is an interesting result, applicable if the lifetime of the KK mode is longer than its tunneling rate. 
It means that, in the large compactification limit,
this KK mode would have a comparable amplitude  in the bulk as the zero mode.
The reason is clear. Although $\Pm$ is peaked in the throat, the throat contributes little to the volume
of the manifold due to the warped measure. As a result of the wave function normalization, $\Pm$ in the bulk should be comparable to that of the graviton. 

In the throat dominated region, however, one finds
\ba
\label{throatratio}
{\cal {R}}_1 \sim {\beta_2}^6 \sim {\beta_1}^6 \sim h_{A}^{6}
\ea
which is very small, as one expects. This just indicates that
the excited modes are confined in the throat.
One can also consider the ratio of the wavefunction at $\tau=\tau_c$ 
for the excited states compared to the zero mode 
 \ba
 \label{ratio}
&& {\cal {R}}_2 =  \left|\frac{\Pm(\tau_c)}{\Psi_{g} } \right|^2
\ea
In the throat dominated region ${\cal R}_2$ scales like $\beta_{13}^6$ which can be very small for small enough $\beta_{1}$. In the bulk
dominated region ${\cal R}_2$ reaches the constant value proportional to  $\beta_{23}^6$,
which is the roots of Eq.(\ref{limit2,2}). In this limit
\ba
{\cal R }_2 \simeq 3, 24, 104
\ea
for $n=1, 2$ and $n=3$ respectively. 
These numbers are constant and are proportional to $x_{3,n}^6$, where $x_{3,n}$ is the $n$-th root of  $J_3$ from Eq.(\ref{limit2,2}). These asymptotic values are independent of choice of $\beta_1$, as long as it is in the bulk dominated regime.

\section{Implication for Reheating}

Although how brane inflation takes place in string theory has been  
studied quite extensively recently, the issue of how inflation ends  
remains a relatively unexplored issue (for early analyses, see e.g.,  
Ref.\cite{Shiu:2002xp,Cline:2002it}). The collision and annihilation  
of a pair of D3-$\D$3-branes will release energy in the form of  
massive closed fundamental strings as well as D-strings. Besides the  
very large ones that may end up as cosmic strings, most of the energy  
will end up as low-lying string modes. For closed strings, the lowest  
mode is the graviton. If too much energy ends up in gravitational  
radiation, this will be incompatible with big bang nucleosynthesis.  
For the brane inflationary scenario to be successful, sufficient  
energy has to be dumped into open string modes, i.e., standard model  
particles. If the standard model branes are sitting at the bottom of  
the inflationary throat, presumably this is not a problem. However,  
if they are sitting in the bulk (say D7-branes wrapping 4 cycles) or  
at the bottom of another throat, then the wave function of the closed  
string modes must be able to tunnel out of the inflationary throat to  
the standard model throat. This is the heating (or
pre-heating/reheating) problem in brane inflation.
Some recent analyses begin to address this issue \cite{Barnaby:2004gg,Kofman:2005yz,Chialva:2005zy,Frey:2005jk}.
It is argued that most of the energy will end up in massive closed  
string
modes \cite{Lambert:2003zr, Chen:2003xq, Leblond:2005km}, which
then decay mostly to closed string KK modes.
It is then important to see how easy it is to transfer the KK energy  
to open strings in another throat.
In this paper, we calculate the tunneling rates of S-wave KK modes  
from the inflationary throat to the bulk and to another throat  
(similar results have been obtained by Chen\footnote{Private discussions.}). Simple scaling argument  
shows that they are suppressed by higher powers of the warp factor  
than in the Randall-Sundrum case. Very similar  properties are  
expected for the KK modes with $S^{3}$ angular momenta while  
qualitatively similar properties are expected for the KK modes with  
$S^{2}$ angular momenta. Due to resonance effect, we find that the  
tunneling rate is very sensitive to the details; but, in any  
realistic scenario, the tunneling of a KK mode from one throat to  
another throat may not be as suppressed as naively thought.
This is encouraging for a successful exit from brane inflation.  
However, the result may be very sensitive to the specific properties  
of the model. Clearly, a more detailed analysis is important.
the above energy spectrum.

\vspace{5mm}

{\bf Acknowledgment}


\vspace{3mm}

We thank Xingan Chen, Jim Cline, Igor Klebanov, Gary Shiu,  Bret Underwood, 
Ira Wasserman and Tung-Mow Yan for valuable discussions. 
We also thank Giacomo Cacciapaglia, Matthew Nobes and Seong-Chan Park
for useful advice on numerical methods.
This work is supported by the National Science Foundation under grant PHY-0355005.

\vspace{3mm}

\appendix

\section{Spherical Harmonics}

One of the technical issues is the geometrical interpretation
of the azimuthal directions of $ds_6^2$ in Eq(\ref{deformedmetric}). 
As mentioned before, at $\tau=0$, $g^5, g^4$ and $g^3$ combine to
form a round $S^3$, while $g^1$ and $g^2$ form a configuration
which is topologically a $S^2$ but shrinks to zero size at $\tau=0$. 
For large $\tau$, on the other hand, these five angular coordinates 
asymptotically matches to a $T^{1,1}$. 
To simplify the calculation, we approximate this situation to an easier 
case where the angular part is $S^2\times S^3$.

The $Q^{(jlm)}$ are harmonics on $S^3$ given by \cite{gerlach}
\ba
Q^{(jlm)}(\chi, \theta, \phi) = F^{jl}(\chi)Y^{(lm)}(\theta, \phi) \\ \nonumber
\ea
where $F^{jl}(\chi)$ are the Fock harmonics,
\ba
F^{jl}(\chi) = (\sin \chi)^l \frac{\partial^{l+1} (\cos j\chi)}{\partial (\cos\chi)^{l+1}}
\ea
Here, $Q^{(jlm)}$ satisfies the following orthogonality condition
\ba
\int d\Omega_3\, Q^{(j\,l'\,m')}\,Q^{(j'\,l''\,m'')}=\delta^{jj'}
\delta^{l'l''}\delta^{m'm''}\, .
\ea
Calling the $S^2$ harmonics $Y^{(l'm')}$, we see that
the wavefunction takes the form
\ba
\Psi({\bf y})=  \sum_{njlml'm'} \psi_{njl'}(r)Q^{(jlm)}(\chi, \theta, \phi)Y^{(l'm')}(\theta ', \phi ')
\ea 

Since $SO(4) \simeq SO(3) \times SO(3)$, we may decompose them into ${\hat Q}_{{\hat j},m,m'}$
instead, where ${\hat j} =0, 1/2, 1, ...$, and $m, m'  = -{\hat j}, . . . ,{\hat j}$. Here $j = 2 {\hat j} +1$, with degeneracy $j^{2}$.

\end{document}